\documentclass[aps,showpacs,nofootinbib]{revtex4}
\usepackage{epsf,latexsym}
\usepackage{amsfonts}
\usepackage{graphicx}

\begin{document}

\title{Minimum-length Ricci scalar for null separated events}
\author{Alessandro Pesci\footnotetext{e-mail: pesci@bo.infn.it}}
\affiliation
{INFN Bologna, Via Irnerio 46, I-40126 Bologna, Italy}

\begin{abstract}
We consider spacetime endowed with a zero-point length,
i.e. with an effective metric structure which allows
for a (quantum-mechanically arising) finite distance $L_0$
between events in the limit
of their coincidence.
%
Restricting attention
to null separated events,
we find an expression for the Ricci (bi)scalar
in this zero-point-length metric;
this is done for when geometric circumstances
are such that the collection of all null geodesics
emerging from a point $P$ has all the information
needed to fix the value of scalar curvature at $P$.   
%
Taking then the coincidence and further $L_0 \to 0$ limits,
we find that this expression does not reduce to the Ricci scalar $R$
of the ordinary metric
but to $(D-1) R_{ab} l^a l^b$ in $D$-dimensional spacetime ($D \ge 4$),
where $R_{ab}$ and $l^a$ are the ordinary Ricci tensor and tangent vector 
to the null geodesics.
This adds nicely
to the existing results for time and space separations.
This finding seems to give further support
to the view that the quantity $R_{ab} l^a l^b$,
ubiquitous in horizon thermodynamics,
embodies
something which remains as a relic/remnant/memory
of a quantum underlying structure for spacetime
in the limit of (actual detectability of) this quantumness
fading away,
and which as such should enter the scene
when aiming to derive/motivate
the
field equations.
Further,
it 
turns out to be
the same quantity used
in an existing derivation of field equations
from a thermodynamic variational principle,
thus adding further evidence of an origin as
quantum-spacetime relic 
for the latter.
\end{abstract}


\maketitle

$ $
\section{Motivation}

The existence of
a deep connection
between gravity and thermodynamics/information theory
is by now
a well-established theoretical fact.
It actually came as a genuine surprise
when, by combining the purely geometric concept 
of a general-relativistic black hole  
with the basic tenets of quantum mechanics, 
it undeniably emerged for the first time \cite{HawD};
indeed, no trace of any thermodynamic feature can be found
in the approach to field equations
based on the extremization 
of the Einstein-Hilbert action
with respect to the metric.
The link with thermodynamics
became
even more evident 
when it was found 
that Einstein's equations
--which we can think of as what axiomatically defines,
together with the equivalence principle, 
general relativity \cite{RinA}--
are of thermodynamic nature themselves, 
a result
(built on Bekenstein's notion
of horizon entropy \cite{BekA, BekB, BekC})
which was first derived in Ref. \cite{JacB}.
This is a very intriguing fact, 
which 
can be seen as raising issues
at a fundamental level, 
for randomness turns out this way to be present
at the heart of general relativity,
and of all of physics with it 
(cf. Ref. \cite{Penrose}).
And yet, any attempt to gain an understanding
of gravity purportedly deeper than that provided by general
relativity should try to provide explanation for this and/or exploit it as a
useful hint.

Beside attempts to construct
full-fledged quantum theories of gravity   
--and potentially to their benefit
(think e.g. of the holographic principle \cite{tHoA, Sus})-- 
there is then clearly a point in the many approaches 
to gravity which try to gain as much insight as possible by
building on thermodynamics.
Among the latter,
some attempts bring to a derivation of field
equations in a thermodynamic setting, 
this clearly implying to go
beyond strict recognition of a thermodynamic meaning for
these equations, breaking thus new ground with respect to
the result \cite{JacB}.    
This seems
to definitely be the case in Refs. \cite{PadG_, PadF, PadN},
where
the field equations of general relativity 
(actually, of a large class of theories of gravity)
have been fully derived
from a thermodynamic variational principle
based on
a functional $S$ representing total (matter+gravitational)
entropy.
This approach has come a long way since then,
up to the point of providing evidence for
dismissing geometry as the primary language
to describe classical gravity, and to be replaced 
entirely by thermodynamic language
\cite{PadK, PadL, Pad05_bis} (see Ref. \cite{Pad20} for a review).

In more recent times
the so-called qmetric has been introduced
as a means to endow spacetime
with a lower limit zero-point length $L_0$ 
(of quantum origin) \cite{KotE, Pad01, KotI}.
It
appears to be a potentially useful tool 
when trying to describe spacetime on small scales
\cite{Pad06, Pad12, Pad05, ChaD}.
The qmetric has turned out to be strictly 
intertwined with the results 
\cite{PadG_, PadF, PadN, PadK, PadL, Pad05_bis};
this is most evident when considering the Ricci scalar $R$. 
Indeed, the expression found for
the qmetric Ricci biscalar $R_{(q)}(p, P)$, 
in the case of space or time separated events $P$ and $p$,
is such that
when we take
the coincidence limit $p \to P$
and further let $L_0 \to 0$,
we get $R_{(q)} \to \varepsilon R_{ab} t^a t^b$ 
($\varepsilon = t^a t_a$,
where $t^a$ is the unit tangent vector to the geodesic
connecting $P$ and $p$) \cite{Pad01, KotI};
now, the point is that this expression
can be shown to reproduce in a specific sense
($R_{(q)}$ is calculated in Riemannian spaces,
then analytically continued from the Euclidean to Lorentz sector)   
(the gravitational part of) the $S$ mentioned 
above, which is proportional to $R_{ab} l^a l^b$ with $l^a$ null
\cite{Pad01, KotI, Pad02},
thus establishing a link 
between two apparently unrelated, 
or not immediately recognizable as related, approaches.

Moreover,
the limiting expression of $R_{(q)}$ 
reported above clearly gives
$\lim_{L_0\to 0} \lim_{p \to P} R_{(q)} \ne R$.
This amounts to saying that
$R$,
namely something
which intrinsically belongs to geometry
and which plays the role of the Lagrangian in the derivation 
of field equations
in the Einsteinian geometric view of gravity,
has in the qmetric (where, of course,
it keeps being of exquisitely geometric nature)   
an expression related to 
$R_{ab} l^a l^b$,
which,
in general relativity complemented with basic tenets of quantum mechanics,
is interpreted as heat density of horizons,
namely something exquisitely of thermodynamic nature.
And this persists also in the $L_0 \to 0$ limit,
when one would naively expect $R_{(q)}$ to become simply $R$.

What this seems to imply is that the
qmetric geometry has an intrinsic thermodynamic flavor,
where the word `intrinsic' here means that
we do not construct quantities which do have a thermodynamic interpretation
but, on the contrary, we compute the expression of well-known
quantities which characterize the geometry and what we get
is thermodynamic expressions for free. 
In addition,
the Lagrangian itself in the Einstein-Hilbert action
becomes in the qmetric
something akin to heat,
and this persists 
in the limit in which the quantum length $L_0$ vanishes.  
The effectiveness of the thermodynamic interpretation of gravity 
would thus appear, according to this,
as a sort of relic of 
an underlying quantum structure for spacetime,
with $R_{(q)}$ seemingly playing some significant role
in this geometry-thermodynamics connection.
Of course the result in the limit above  
refers, strictly speaking, directly to $R_{ab} t^a t^b$ 
and not $R_{ab} l^a l^b$, which is what instead 
the thermodynamics of horizons demands.
However one of the aims of the present study is
to try to address this point by considering
the qmetric and the Ricci scalar in it through null separated events.
What seems to happen indeed,
is, as already anticipated in the Abstract,
that the limit above turns out to be in this case
proportional to $R_{ab} l^a l^b$.

On the side of the attempts 
aiming at constructing a complete quantum description 
of gravity,
a proposal has been recently made,
within causal dynamical triangulation,
of a new geometric observable
capable of playing the role 
of a quantum Ricci curvature \cite{LolA, LolB}.
This might give the opportunity 
to cross compare
among very
different approaches to quantum curvature
(one proceeding from nonsmooth metric spaces, 
the other (the qmetric) trying to extend
a smooth metric down to the quantum limit)   
and provides further motivation for an 
as complete as possible account of curvature 
on the side of the qmetric.   

In view of these points,
in the present work we  
try to extend
the investigation 
of the qmetric biscalar $R_{(q)}$,
in particular seeking, as mentioned, to find an expression for it
(or at least approximating it)
for null separated events,
a task that is complicated a little
by the need to handle 
--when writing down the qmetric transverse
to the geodesics connecting the events--
both the tangent vector and
the auxiliary (null) vector. 
After obtaining such an expression,
we then proceed to
explore
its coincidence
and further $L_0 \to 0$ limits.

\section{Gauss--Codazzi relations for null equi-geodesic hypersurfaces}

To evaluate the zero-point-length Ricci scalar for null separated 
events, we exploit the Gauss--Codazzi framework conveniently generalized 
to the case of null hypersurfaces. This --taking a cue 
from the approach of Refs. \cite{KotG, KotI} 
for timelike/spacelike separated events--
with the aim of having
an expression for the Ricci scalar
in which the components of the Riemann
and Ricci tensor do not appear, 
a fact this beneficial in view of the qmetric calculation.
Given a generic point $P$ in a 
$D$-dimensional spacetime $M$ ($D \ge 4$),
we introduce an arbitrarily chosen, 
freely falling observer $\hat V^a$ at $P$,
whose aim is to ideally get information on the geometry of $M$  
through null geodesics from $P$.
We consider a congruence of 
future-directed null geodesics $\gamma$ emanating from $P$,
affinely parametrized through distance 
according to the observer $\hat V^a$.
After proceeding to extract information
from these geodesics,
our goal is to see what we can say about curvature
according to the null qmetric \cite{PesN} associated with this congruence.

The chosen parametrization 
$\lambda = \lambda(p, P)$ ($\lambda(P, P) = 0$)
gives
$l^a {\hat V_a} = -1$ at $P$ for each geodesic,
where
$l^a= dx^a/d\lambda$ 
is the tangent vector to the considered geodesic,
where $\{x^a\}$,  $a = 1, ..., D$, are coordinates of $p$
[our conventions are that
the metric $g_{ab}$ has 
signature $(-, +, +, +, ...)$, 
and the Riemann tensor is defined by
$[\nabla_c, \nabla_d] \, v^a = {R^a}_{bcd} \, v^b $, $v^a$ vector].
At any point of each of these geodesics
we introduce an observer 
with velocity $V^a$ given
by the parallel transport of ${\hat V^a}$
along the geodesic connecting $P$ and $p$
($V^a = {\hat V^a}$ at $P$).
$V^a$ is
such that
$l_a V^a = -1$
at every point of these geodesics.
$\lambda$ is thought of as the distance
along any one of the geodesics as measured by observers with 
(parallel-transported)
velocity $V^a$; in a local frame at $p$
with velocity $V^a$, $\lambda$ is the 
spatial distance from $p$.

Let us call $L$ the $(D-1)$-dimensional submanifold swept
by all the geodesics of the congruence,
i.e.
\begin{eqnarray}
\nonumber
L = \big\{p \in M: 
\ {\rm quadratic \ distance} \
\sigma^2(p, P)=0, \ {\rm and} \ p \ {\rm is \ in \ the \ future \ of} 
\ P\big\}. 
\end{eqnarray}
It has $l^a$ as normal vector
(since, from 
$0 = \frac{d\sigma^2}{d\lambda} 
= \frac{dx^a}{d\lambda} \, \partial_a \sigma^2$,
$l_a$ must be parallel to the
null normal $\partial_a \sigma^2$)
and can be locally regarded as a collection
of spacelike $(D-2)$-surfaces
$\Sigma(P, \lambda)$,
i.e.
$L$ is locally
$L = \gamma \times \Sigma(P, \lambda)$,
with
\begin{eqnarray}
\nonumber
\Sigma(P, \lambda) =
\big\{p \in L: 
\ \lambda(p, P) = \lambda \, (> 0)
\big\},
\end{eqnarray}
with $\lambda$ fixed.
For $L$,
we can then use coordinates
$(\lambda, \theta^A)$ where $\theta^A$, $A = 1, ..., D-2$,
label the different geodesics $\gamma$.
%
We can be sure that this construction is consistent 
provided we limit ourselves to points $p$ 
such that $\lambda(p, P)$ is small enough
that no focal points develop in the congruence (in addition to
the starting point $P$) (for on the contrary any such focal point 
would necessarily have multiple
assignments of $\theta^A$).

Introducing
$m^a = dx^a/d\nu = 2 V^a - l^a$ 
as an auxiliary null vector at $p$
[then parallel-transported along any $\gamma$, 
and with $m_a \, e^a_A = 0$ 
(with $e^a_A \equiv \partial x^a/\partial \theta^A)$ 
and $m_a l^a = -2$
at every $p \in L$],
the induced metric on $\Sigma(P, \lambda)$ can be expressed as 
(cf. e.g. Ref. \cite{PoiA})

\begin{eqnarray}\label{62_1}
h_{ab} = g_{ab} + \frac{1}{2} \, l_a m_b + \frac{1}{2} \, m_a l_b, 
\end{eqnarray}
with ${h^a}_b$ acting as a projector into the space $T(\Sigma)$ tangent
to $\Sigma(P, \lambda)$,
${h^a}_b \, l^b = 0 = {h^a}_b \, m^b$ and  
${h^a}_b \, {h^b}_c = {h^a}_c$.

In Ref. \cite{Gem} a relation connecting 
the Riemann
tensor ${{R_\Sigma}^{\, a}}_{bcd}$ intrinsic to null hypersurfaces
with the Riemann tensor ${R^a}_{bcd}$ of $M$
was given (a generalization to null hypersurfaces of 
the so-called first Gauss--Codazzi identity), 
which we take as our starting point.
In our circumstances,
it reads

\begin{eqnarray}\label{63_3}
{R_\Sigma}_{\, abcd} 
=
R_{efgh} \, {h^e}_a {h^f}_b {h^g}_c {h^h}_d
+ K^{(r)}_{bd} K^{(r)}_{ac} - K^{(r)}_{ad} K^{(r)}_{bc}
- K^{(V)}_{bd} K^{(V)}_{ac} + K^{(V)}_{ad} K^{(V)}_{bc},
\end{eqnarray}
where 
\begin{eqnarray}
\nonumber
K^{(r)}_{ab} &\equiv& {h^c}_a {h^d}_b \nabla_c r_d \\
\nonumber
K^{(V)}_{ab} &\equiv& {h^c}_a {h^d}_b \nabla_c V_d
\end{eqnarray}
[cf. Eq. (50) of Ref. \cite{Gem}],
with
$
r^a \equiv l^a - V^a,
$
and $V^a$,  defined as above, extended outside $L$.
Due to the integrability of $V^a$ at any point $p$ of $L$
[$V^a$ is orthogonal to the hypersurface
$i_{(r)} \times \Sigma(P, \lambda)$,
where $i_{(r)}$ is the integral curve of $r^a$ through $p$],
${{R_\Sigma}^{\, a}}_{bcd}$ in the relation above
is defined by
\begin{eqnarray}
\nonumber
{Y^a}_{\| cd} - {Y^a}_{\| dc} 
=
- {{R_\Sigma}^{\, a}}_{bcd} Y^b
\end{eqnarray}
for any $Y^a \in T(\Sigma)$
[cf. Eq. (55) in Ref. \cite{Gem}],
where a double stroke means differentiation 
with respect to the connection relative to the induced metric. 

A little algebra
permits to express Eq. (\ref{63_3}) in terms
of the vectors $l^a$, $m^a$, 
defining the induced metric in Eq. (\ref{62_1}),
as 

\begin{eqnarray}\label{65_1}
{R_\Sigma}_{\, abcd} 
=
R_{efgh} \, {h^e}_a {h^f}_b {h^g}_c {h^h}_d + \frac{1}{2} \Big(
- K_{bd} {\bar K}_{ac} + K_{ad} {\bar K}_{bc}
- {\bar K}_{bd} K_{ac} + {\bar K}_{ad} K_{bc}\Big),
\end{eqnarray}
where
\begin{eqnarray}
\nonumber
K_{ab} &\equiv& {h^c}_a {h^d}_b \nabla_c l_d 
= K_{ab}^{(V)} + K_{ab}^{(r)} \\
\nonumber
{\bar K}_{ab} &\equiv& {h^c}_a {h^d}_b \nabla_c m_d
= K_{ab}^{(V)} - K_{ab}^{(r)}.
\end{eqnarray}
Equation (\ref{65_1}) does coincide with that provided,
by other means and in complete generality,
in Ref. \cite{ChaC_bis}
[Eq. (99) there].
From it, we get in particular the scalar relation

\begin{eqnarray}\label{106_3}
R_\Sigma 
=
{R^e}_{fgh} \, {h^g}_e h^{fh} - K {\bar K} + K^{ab} {\bar K}_{ab},
\end{eqnarray}
$K = {K^a}_a$, ${\bar K} = {{\bar K}^a}_{\; \> \, a}$,
where we used the symmetry of $K_{ab}$
(from the hypersurface orthogonality of $l^a$).
The connection with the Ricci scalar $R$ is made
using the easily algebraically established relation
\begin{eqnarray}
\nonumber
{R^e}_{fgh} \, {h^g}_e h^{fh}
=
R + 2 R_{ab} \, l^a m^b + \frac{1}{2} R_{efgh} \, l^e m^f m^g l^h
\end{eqnarray}
(where $R_{ab}$ is the Ricci tensor).
Equation (\ref{106_3}) then becomes the
relation

\begin{eqnarray}\label{106_4}
R_\Sigma 
=
R + 2 R_{ab} \, l^a m^b + \frac{1}{2} R_{efgh} \, l^e m^f m^g l^h
- K {\bar K} + K^{ab} {\bar K}_{ab}.
\end{eqnarray}

As it stands, this relation is not very useful
in view of deriving the qmetric expression
of the Ricci scalar.
$R$ is indeed explicitly expressed in terms of the Ricci and Riemann tensors'
components,
and, when going to the qmetric setup,
this would imply being
able to obtain $R_{(q)}$ only if we had general expressions for
$R_{ab}^{(q)}$ and $R_{abcd}^{(q)}$,
meaning trying to solve our problem with a bigger one.
This is at variance with what happens for
timelike or spacelike congruences from $P$,
for which we have a relation
in place of Eq. (\ref{106_4}) with no explicit dependence
on the Ricci and Riemann tensors' components in it,
and this can be usefully exploited to derive $R_{(q)}$
for these cases [see Eq. (16) of Ref. \cite{KotG}].

One specificity of our situation here
is that the submanifold $L$ swept by the geodesics
of the congruence is $(D-1)$ (not $D$) dimensional.
%
All our congruence-related quantities are thus defined
only on a submanifold $L$ with one dimension less
than the whole spacetime.

To see what we can do,
we start by noticing that
any couple of events $p'$ and $p''$ near $p$, 
the first along $\gamma'$ (with tangent $m^a$) from $p$ and
the second along $\gamma$,
which are taken as simultaneous according to the observer $V^a$ at $p$,
give

\begin{eqnarray}\label{dnu_dlambda}
d\nu(p', p) = d\lambda(p'', p), 
\end{eqnarray} 
where
$d\nu(p', p) \equiv \nu(p') - \nu(p) = \nu(p')$
taking $\nu(p) = 0$,
and
$d\lambda(p'', p) \equiv \lambda(p'',P) -\lambda(p, P).$

Let us consider first a static metric and
a situation in it in which $V^a$ at $p \in L$ is directed
as the timelike Killing vector field $\xi^a$ associated 
with the staticity of the metric.
$V^a$ is then orthogonal to a spacelike hypersurface $S$ which acts
as a symmetry surface for spacetime.
In this circumstance,
$\gamma'$ brings back from $p(\lambda)$ to nearby points $p'(\nu)$ 
at same
spatial coordinates on $S$ as points $p(\bar\lambda)$,
where $\bar\lambda = \lambda - \nu$
[i.e. points reached along $\gamma$ from $P$, before $p$],
but at later times.
Since spacetime is static
and this time delay then has no effect, 
this amounts to saying that, at $p$,
if the quantity we consider can be thought of as having no dependence on time 
by its own,
its differentiation along $m^a$ ought to be taken as the opposite
of differentiation along $l^a$.
%
%
%
%
Specifically,
the vector fields
$l^a$, $m^a$ and the $\theta^a_A$'s
can be thought of as
parallel transported along $V^a$,
and

\begin{eqnarray}\label{time}
V^a \nabla_{a} {\cal Q} = 0  
\end{eqnarray}
with them at $p$,
for any quantity $\cal Q$,
be it a scalar, vector or tensor,
specified at any point of $L$
and assigned with the congruence,
i.e. which does not have its own life
independent of the congruence 
and is defined by the fields above as evaluated on $L$
[for example, the vector $z^a = l^b \nabla_b l^a (= 0)$
or the tensor $y_{ab} = {h^c}_a {h^d}_b \nabla_c l_d (= K_{ab})$].
For any such quantity, we then get
(from $m^a = - l^a + 2 V^a$)

\begin{eqnarray}\label{Q}
m^a \nabla_a {\cal Q} =
- l^a \nabla_a {\cal Q}  
\end{eqnarray}
at $p$.

These relations look very promising in view of further
manipulation (and simplification) of Eq. (\ref{106_4})
(as we will see in a moment).
We realize however that these results 
are not due to the extension we have chosen off $L$
but to this extension {\it and} the fact we have taken spacetime
as being static (with respect to time as provided by $V^a$).
Indeed,
let us consider for example
the vector $l^b \nabla_b l^c$.
We have

\begin{eqnarray}\label{Q1_28_1}
V^a \nabla_a(l^b \nabla_b l^c)
&=&
(V^a \nabla_a l^b) \nabla_b l^c + V^a l^b \nabla_a \nabla_b l^c
\nonumber \\
&=&
V^a l^b \nabla_a \nabla_b l^c
\nonumber \\
&=&
V^a l^b \nabla_b \nabla_a l^c - {R^c}_{dba} l^d l^b V^a
\nonumber \\
&=&
- {R^c}_{dba} l^d l^b V^a,
\end{eqnarray}
where
the second equality stems from the parallel transport of $l^b$ along $V^a$,
the third from
$\nabla_a \nabla_b l^c - \nabla_b \nabla_a l^c = - {R^c}_{dba} l^d$,
and the fourth from the parallel transport of $V^a$ along $l^b$ and again
the parallel transport of $l^c$ along $V^a$.
We see here that the possibility that
$V^a \nabla_a(l^b \nabla_b l^c)$ vanishes
is in any case linked to some specific symmetries
the Riemann tensor has to possess 
at points of $L$,
a requirement which adds
to how we chose to extend $l^a$ off $L$.
These are just the symmetries the Riemann tensor has
thanks to the fact that spacetime is static.

For a generic spacetime,
there is then no hope
of obtaining that Eqs. (\ref{time}) and (\ref{Q}) hold true.
Indeed, whichever way we may think to extend off $L$
the vectors
tangent to $L$,
we certainly cannot
change this way the properties of ${R^a}_{bcd}$ at points of $L$.
This seems to definitely
close in the negative
the issue of finding a qmetric-amenable form
of Eq. (\ref{106_4}),
and the possibility with it of finding
an expression for $R_{(q)}$.

These considerations point to the root
of the difficulty, namely that 
for generic $M$, our hope of expressing
$R$ at $P$ in terms of quantities relative to $L$
does not work.
We may ask:
what if we deform the spacetime $M$, not on points of $L$,
but only outside $L$, to a new spacetime $M^\sharp$
in search of more favorable geometric circumstances,
making sure at the same time
that our congruence keeps being a (null) congruence?

${R^a}_{bcd}$ changes
on points of $L$,
for, even if we do not touch $L$, we actually do touch
near $L$.
One might have the hope however 
that perhaps the Ricci scalar at $P$
does not change.
It is true indeed that
in general the Ricci scalar
is given at a point 
with a congruence emerging from that point 
[it is readable in the ratio (in the shrinking limit) 
between the area of the equigeodesic surface of the congruence
emerging from that point
in the actual spacetime under consideration and the area we would have
in Minkowski spacetime]
and
this implies that, given a congruence sweeping spacetime 
in a neighborhood of $P$, it has in it all the information regarding
$R$ at $P$.
Our congruence however,
is short of one dimension
with respect to what would be needed to probe the whole spacetime
at $P$. 
It is then impossible in general to have on $L$ all the information
needed to get $R$ at $P$.

In other words,
to derive $R$ at $P$ on the basis of a congruence of null geodesics 
emerging from it is in general
intrinsically impossible, for these do not sweep the whole
spacetime around $P$.
This same thing however,
suggests what are the geometric conditions
which allow for our program to be accomplished.
When geometric circumstances are indeed such that $L$
does determine $R$ at $P$, it does become meaningful
to provide an expression for $R$ on the basis of the sole
quantities defined for $L$, i.e. in terms of null geodesics.
Starting from the given spacetime $M$ we can then try to deform it
to a new spacetime $M^\sharp$ in which these geometric circumstances 
are indeed realized.
Our aim becomes then
to cover in this way all spacetimes which do admit a description
of the Ricci scalar at a given point in terms of hypersurfaces swept by null
geodesics from that point. 

But,
can we be sure that spacetime can be deformed
respecting the prescriptions
just mentioned?
Here is where what we learned
in the case of static spacetime
can help.
We choose as $M^\sharp$
a spacetime which,
indefinitely close to $L$
(i.e. ``instantaneously''),
consists
of $L$ parallel transported along $V^a$
(then, any point of $L$ as well as any
scalar, vector or tensor
assigned with the congruence
in the meaning above)
according to the covariant derivative in $M$ taken,
at any point of $L$, as defining the covariant derivative in $M^\sharp$.
We have thus in one stroke
both a new spacetime $M^\sharp$
and a chosen extension off $L$ of the vectors
$l^a$, $m^a$ and the $\theta^a_A$'s.
This,
i) can always be done,
ii) guarantees that the $\gamma$'s are also geodesics in the new spacetime
(they indeed still satisfy $l^b \nabla_b l^a = 0$),
iii) implies that,
if the vector fields $l^a$, $m^a$ and the $\theta^a_A$s
are parallel extended off $L$ in $M$, 
the covariant derivatives of the same in any direction
do coincide in $M$ and $M^\sharp$, 
and iv) $L$ is left unchanged, and in $M^\sharp$ any quantity
defined on $L$ is parallel transported along $V^a$.
What we do at the end,
is to consider a sort of instantaneously static spacetime ``tangent''
to the original spacetime at the common
submanifold $L$.

As it concerns the qmetric spacetime,
it happens, as we will see next
(and can also be seen in Ref. \cite{PesN}), 
that the submanifold $L$
is also the submanifold $L_{(q)}$
swept by the qmetric null geodesics,
so it is left untouched by the deformation, and then it is common
to the qmetric versions of the two spacetimes.
That same q-quantities defined on $L_{(q)}$ 
can then be used to express the qmetric Ricci scalar
of that qmetric spacetime, any time it exists, 
in which the (qmetric) Ricci scalar at a point
is fully determined by the (qmetric) null geodesics at that point.
This qmetric spacetime does not need to coincide
with the qmetric version $M^\sharp_{(q)}$ of $M^\sharp$.


From now on, our ordinary spacetime will be $M^\sharp$.
%
This spacetime shares with spacetime $M$ 
(endowed with parallel extensions of the fields
$l^a$, $m^a$, and the $\theta^a_A$'s off $L$)
the submanifold $L$,
as well as the tangent vectors $l^a$, $m^a$, and the $\theta^a_A$'s
at any point of $L$ and the covariant derivatives
(in any direction) at any point of $L$ of these same vectors;
it also shares the derivatives of derivatives of vectors,
if they are taken along $L$ ($L$ is left untouched;
everything that happens ``along'' $L$ remains the same).
It has then to share
the metric tensor
at any point of $L$ (scalar products of any couple of vectors
are the same in $M$ and $M^\sharp$ at any point of $L$)
as well as the connection
and the (ordinary) derivatives of the metric
(the covariant derivatives of the vectors
$l^a$, $m^a$, and the $\theta^a_A$'s in any direction 
are the same in $M^\sharp$ and $M$). 
$M^\sharp$ then also shares with $M$
$K_{ab}$, $\bar K_{ab}$ and $R_{\Sigma}$ at any point of $L$.
Quantities without the $^\sharp$ symbol
are evaluated in $M$;  
but, to stress the facts just mentioned
and to not overload notation,
even if we work entirely in $M^\sharp$
we shall relegate the use of the $^\sharp$ symbol
only to quantities
which differ, or might differ, in $M^\sharp$ and $M$.

Since
the qmetric along any geodesic of the congruence
depends only on the geodesic itself \cite{PesN},  
the two 
spacetimes are in this respect
completely equivalent,
in the sense that any quantity
in the expression
of the qmetric along
any one of these geodesics
is identical in $M$ and $M^\sharp$
(as we will consider in more detail later).

Terms which involve explicitly $R_{ab}$ and $R_{abcd}$ are
different in the two spacetimes 
also
at the points of L
(in addition to the points off $L$);
terms of this kind however, as we will see in a moment,
simply do disappear (in general without this requiring 
all of them to vanish)
from Eq. (\ref{106_4}) in $M^\sharp$
(this is the reason why we went to $M^\sharp$ in the first place).
As for
the Ricci scalar,
we know it is different
in the two spacetimes; 
it is $R^\sharp$ which lends itself to being expressed
in terms of quantities defined for $L$ alone.

The end result is that Eq. (\ref{106_4}),
when considered in $M^\sharp$,
becomes an equation which still involves quantities
in $M$, but without any presence left
of the unwanted curvature terms,
and with reference
to $R^\sharp$ in place of $R$.
Let us see how this happens.

In $M^\sharp$ we have
that the $^\sharp$ version of Eq. (\ref{time}) holds true,
by construction.
We then also have
the $^\sharp$ version of the relations (\ref{Q}).
Choosing as $\cal Q$ a scalar or a vector,
the latter can be written as

\begin{eqnarray}\label{133_0}
(m^a \partial_a \Phi)^\sharp 
=
- l^a  \partial_a \Phi
\end{eqnarray}
and

\begin{eqnarray}\label{133_0_bis}
(m^a \nabla_a z^b)^\sharp
=
- l^a  \nabla_a z^b
\end{eqnarray}
at $p \in L$
for any scalar $\Phi$ and any vector $z^b$
assigned with the congruence
[in the specific sense described below
Eq. (\ref{time})].
In these expressions,
when in the definition
of $\Phi$ or $z^a$ do enter simply the vectors
$l^a$, $m^a$ and the $\theta^a_A$'s,
and not e.g. their derivatives,
the $^\sharp$ mark is
not required.

When in particular
$\Phi = {\rm const}$ along $\gamma$,
we get

\begin{eqnarray}\label{101_1}
(m^a \partial_a \Phi)^\sharp 
= 0
\end{eqnarray}
and,
when $z^a$ is parallel trasported along $\gamma$,

\begin{eqnarray}\label{102_6}
(m^a \nabla_a z^b)^\sharp 
= 0,
\end{eqnarray}
both at $p \in L$.
When $\Phi = {\rm const}$ all over $L$,
Eq. (\ref{101_1}) gives

\begin{eqnarray}\label{100_Phi}
(\partial_a \Phi)^\sharp = 0
\end{eqnarray}
at $p \in L$;
in particular

\begin{eqnarray}\label{100_3}
l^b \nabla_a \, l_b = 0 
\end{eqnarray}
(cf. Ref. \cite{PesN}) and

\begin{eqnarray}\label{100_6}
m^b \nabla_a \, m_b = 0 
\end{eqnarray}
at $p \in L$,
upon choosing
$\Phi = l^b l_b$ or
$\Phi = m^b m_b$
respectively.

In view of this,
specifically upon using Eqs.
(\ref{101_1}) and (\ref{102_6}),
one can verify that on the rhs of Eq. (\ref{106_4})
as written in $M^\sharp$,
the third term vanishes
i.e. 

\begin{eqnarray}\label{107_4}
R_{efgh}^\sharp \, l^e m^f m^g l^h = 0;
\end{eqnarray}
moreover,
the term involving the Ricci tensor 
can be expressed 
entirely in terms of $l^a$, $m^a$, $K_{ab}$, ${\bar K}_{ab}$
as  

\begin{eqnarray}\label{111_0}
2 R_{ab}^\sharp \, l^a m^b
=
- (m^a \nabla_a K)^\sharp - l^a \nabla_a {\bar K}
+ \big[\nabla_a \big(m^b \nabla_b l^a\big)\big]^\sharp 
+ \big[\nabla_a \big(l^b \nabla_b m^a\big)\big]^\sharp 
- 2 K^{ab} {\bar K}_{ab}.
\end{eqnarray}
All of this is spelled out in Appendix \ref{AppA}. 
Equation (\ref{106_4}) can then be given the form

\begin{eqnarray}\label{111_2}
\nonumber
R_\Sigma 
=
R^\sharp - (m^a \nabla_a K)^\sharp - l^a \nabla_a {\bar K}
+ \big[\nabla_a \big(m^b \nabla_b l^a\big)\big]^\sharp 
+ \big[\nabla_a \big(l^b \nabla_b m^a\big)\big]^\sharp
- K {\bar K} - K^{ab} {\bar K}_{ab}.
\end{eqnarray}

We now use the fact that 
the vectors
$m^b \nabla_b l^a$ and $l^b \nabla_b m^a$ are both identically
vanishing on $L$ [the first in view of Eq. (\ref{102_6}), and the second by
construction].
From this, 
and from Eq. (\ref{102_6})
as applied alternately to
$z^a = m^b \nabla_b l^a$ and $z^a = l^b \nabla_b m^a$,
we get

\begin{eqnarray}
\nonumber
\big[\nabla_c \big(m^b \nabla_b l^a\big)\big]^\sharp = 0
\end{eqnarray} 
and

\begin{eqnarray}
\nonumber
\big[\nabla_c \big(l^b \nabla_b m^a\big)\big]^\sharp = 0
\end{eqnarray}
at $p \in L$.

This gives

\begin{eqnarray}\label{112_4bis}
R_\Sigma 
&=&
R^\sharp - (m^a \partial_a K)^\sharp - l^a \partial_a {\bar K}
- K {\bar K} - K^{ab} {\bar K}_{ab},
\nonumber \\
&=&
R^\sharp - l^a \partial_a ({\bar K} - K) 
- K {\bar K} - K^{ab} {\bar K}_{ab}
\end{eqnarray}
[second equality from Eq. (\ref{133_0})]
or

\begin{eqnarray}\label{112_4}
R^\sharp
&=&
R_\Sigma + K {\bar K} + K^{ab} {\bar K}_{ab}
+ (m^a \partial_a K)^\sharp + l^a \partial_a {\bar K}
\nonumber \\
&=&
R_\Sigma + K {\bar K} + K^{ab} {\bar K}_{ab}
+ l^a \partial_a ({\bar K} - K)
\end{eqnarray}
at $p \in L$.
This relation is (kind of) the direct expression of the Ricci scalar
(meaning without any resort to components of Ricci or Riemann tensors)
we were looking for.
%
What we have here (at $p$) is
$R^\sharp$ completely expressed
in terms of quantities defined in terms of $L$
only.

Looking back to the path we followed,
what we have found is that,
given a generic spacetime
and a null geodesic congruence emerging from a point $P$ of it
with a swept $(D-1)$-submanifold $L$,
the Ricci scalar $R^\sharp$ at $p\in L$
in an associated manifold $M^\sharp$ with the same $L$ 
and with $R^\sharp$ completely determined by $L$
is given by
the rhs of Eq. (\ref{112_4}).
%
This, 
thing which is the key point,
without any assumption concerning
the null fields $l^a$ and $m^a$ describing the congruence,
besides
$l^a$ being tangent to the congruence,
having $m^a l_a = -2$,
and both vectors being
parallel transported along the congruence.
%
It does not matter which way the vectors 
of the congruence may be thought of as extended off $L$ in $M$;
in Eq. (\ref{112_4}) indeed (second equality), every quantity is given with $L$
and, if differentiated, this is along $L$.

For later use,
we note
a property
of Eq. (\ref{112_4})
just from a mathematical point of view.
We ask what would happen
if
that same induced metric $h_{ab}$ of Eq. (\ref{62_1})
were given in terms of a 
vector $\hat m^a$
in place of $m^a$,
parallel to it,
and parallel transported along $\gamma$
(in place of $m^a$).
That is if it were

\begin{eqnarray}\label{179_5}
h_{ab} = g_{ab} + \frac{1}{2} \, l_a {\hat m_b} + \frac{1}{2} \, {\hat m_a} l_b
\end{eqnarray}
with
${\hat m_a} e^a_A = 0$, ${\hat m_a} l^a = -2$ and $\hat m^a$ parallel 
transported along $\gamma$,
but with

\begin{eqnarray}\label{mhat}
{\hat m^a} = \frac{dx^a}{d{\hat \nu}}  
= \frac{1}{\mu(\lambda)} \, m^a,
\end{eqnarray}
where $\mu(\lambda)$ is a scalar
[thus with

\begin{eqnarray}
d{\hat \nu(p', p)} = \mu \, d\nu(p', p) = \mu \, d\lambda(p'', p)
\end{eqnarray}
from Eq. (\ref{dnu_dlambda})
($d{\hat \nu(p', p)} \equiv {\hat\nu}(p') - {\hat\nu}(p)$),
for the events $p'$ and $p''$ considered above].

Clearly if
$h_{ab}$ in Eq. (\ref{62_1}) is given by $m^a$
it cannot be given also, with the same numerical coefficients,
by $\hat m^a$.
What we maintain however,
is that it still has a definite meaning to ask
what would have been the form of Eq. (\ref{112_4}),
had the $m^a$ of Eq. (\ref{62_1}) (with $h_{ab}$ and $l^a$ held fixed, assigned)
turned out to be $\frac{1}{\mu}$ times what it actually is.

$R^\sharp$, $R_\Sigma$, $K_{ab}$ and $K$ would have remained unaffected
(the manifold, the submanifold and the vector $l^a$
would have been the same),
while,
\begin{eqnarray}\label{hat_133_0}
({\hat m}^a \partial_a \Phi)^\sharp 
=
- \frac{1}{\mu} \, l^a  \partial_a \Phi
\end{eqnarray}
would have replaced
Eq. (\ref{133_0}) at $p \in L$
[from Eq. (\ref{mhat}), with $l^a$ remaining the same]
for any scalar $\Phi$ contemplated in Eq. (\ref{133_0}).

The relation between $R^\sharp$ and $R_\Sigma$ would 
clearly have remained the same
as in Eq. (\ref{112_4}),
with ${\bar K}_{ab}$ given by

\begin{eqnarray}\label{179_6}
{\bar K}_{ab} 
&=&
{h^c}_a {h^d}_b \nabla_c m_d
\nonumber \\
&=&
{h^c}_a {h^d}_b \nabla_c \big(\mu \, {\hat m_d}\big)
\nonumber \\
&=&
\mu \, {h^c}_a {h^d}_b \nabla_c {\hat m_d}
\nonumber \\
&=&
\mu \, {\hat K_{ab}}
\end{eqnarray}
\big(${\hat K_{ab}} \equiv {h^c}_a {h^d}_b \nabla_c {\hat m_d}\big)$
when expressed in terms of $\hat m^a$,
with the third equality coming from ${h^d}_b \, {\hat m_d} = 0$.  
Equation (\ref{112_4}) would then read

\begin{eqnarray}\label{179_8}
R^\sharp
&=&
R_\Sigma + \mu \, K {\hat K} + \mu \, K^{ab} {\hat K}_{ab}
+ \mu \, ({\hat m^a} \partial_a K)^\sharp + l^a \partial_a \big(\mu {\hat K}\big)
\nonumber \\
&=&
R_\Sigma + \mu \, K {\hat K} + \mu \, K^{ab} {\hat K}_{ab}
+ l^a \partial_a \big(\mu {\hat K} - K\big)
\end{eqnarray}
(${\hat K} = {\hat {K^a}_a}$),
where the second equality is from Eq. (\ref{hat_133_0}).

In other words,
what one learns here is that,
if, by some inscrutable reason,
the derivatives at $p \in L$ 
of mentioned scalars
along the auxiliary vector
were related to the derivatives of the same along 
the tangent vector
to the congruence 
by Eq. (\ref{hat_133_0})
in place of Eq. (\ref{133_0}),
then
the Ricci scalar $R^\sharp$ at $p$
would be given by Eq. (\ref{179_8})
in place of Eq. (\ref{112_4}).


\section{Scalar Gauss--Codazzi relation for null equi-geodesic 
hypersurfaces with zero-point length}

Having found potentially convenient expressions 
for the Ricci scalar on null
equigeodesic hypersurfaces for an ordinary metric, 
we can proceed now to try to find out how to translate them 
into the context of the qmetric $q_{ab}$.
Our hope is to extract this way a formula useful to infer
the zero-point-length Ricci scalar at a point
from the coincidence limit of null separated events.
To this aim,
our first task is to explore what our geometric construction
in a spacetime with an ordinary metric $g_{ab}$
looks like in the same spacetime but with reference to $q_{ab}$.

Let us first recall how the qmetric is defined.
The qmetric
$q_{ab}(x, x')$ is constructed
as a tensorial quantity  which is a function 
of two spacetime points $x$, $x'$
(i.e. it is a bitensor; see e.g. Ref. \cite{PPV}),
where we may think of $x'$ 
as a fixed, ``base'' point, and of $x$, with coordinates $x^a$, 
as a variable point.
$q_{ab}(x, x')$ is regarded
as a second rank tensor at $x$ 
(concerning any biquantity we shall consider in the paper,  
indices are meant to refer to point $x$). 
Its specific expression stems 
\cite{KotE, KotG, KotI}
from requiring 
it to give a new quadratic interval $S$
between $x'$ and $x$
which is assumed to be function of $\sigma^2$ alone
(where $\sigma^2 = \sigma^2(x, x')$ is the quadratic interval 
for $g_{ab}$),
i.e. $S = S(x, x') = S(\sigma^2)$,
with the properties
$i) \, S \simeq \sigma^2$ 
for large separations
and,
in the case of space- or time-separated events, 
$ii) \, S \to \varepsilon \, {L_0}^2$ in the coincidence limit $x \to x'$ 
($\varepsilon = +1$ or $\varepsilon = -1$
for spacelike-connected or timelike-connected events, respectively)
along the geodesic connecting $x'$ and $x$.
Here the parameter $L_0$ is a fundamental length,
whose value is assumed to be determined by the
underlying microscopic quantum properties of matter and geometry. 

For large separations, 
$S(x, x')$ is $\sigma^2(x, x')$,
that is $2 \, Sy(x, x')$, where $Sy(x, x')$ is the Synge world function.
When $x$ approaches $x'$ however, $S(x, x')$ wildly differs
from $2 \, Sy(x, x')$, in that it approaches a finite limit,
while $Sy(x, x') \to 0$. 
This makes it clear that the qmetric $q_{ab}$ 
cannot be considered a metric
in the ordinary sense, 
for, contrary to any such metric, 
it does not give
a vanishing distance in the coincidence limit.

In the case of null separated events
(which clearly give $\sigma^2 = 0 = S$),
we cannot resort to $S=S(\sigma^2)$ to define the qmetric.
It is clear however that,
going to a local frame at $x'$ such that
the given affine parameter $\lambda$ of the geodesic
is in it length and time,
the existence of a limit value $L_0$ for both
reflects in a limit value $L_0$ for $\lambda$.

Then,
writing $\lambda = \lambda(x, x')$ 
for the affine interval 
between $x'$ and $x$,
what one requires is that
$q_{ab}$ gives rise \cite{PesN}
to a new qmetric-affine parametrization
with interval $\tilde \lambda$ between $x'$ and $x$
which must result in a function of $\lambda$ alone,
i.e.
$\tilde \lambda = \tilde\lambda(x, x') = \tilde\lambda(\lambda)$,
and must have the properties
$i)'\, \tilde\lambda \simeq \lambda$ for large separations
(meaning when $\lambda \gg L_0$)
and $ii)' \, \tilde\lambda \to L_0$ in the coincidence limit 
$x \to x'$ 
along the null geodesic
(i.e. when $\lambda \to 0$). 
Properties $i)'$ and $ii)'$ reflect
for null geodesics
what is implied
by properties $i)$ and $ii)$ for spacelike/timelike geodesics.

Compatibility with $q_{ab}$, uniquely fixes the qmetric connection
${{\Gamma^a}_{bc}}^{(q)}$ in terms of $q_{ab}$ (and ${\Gamma^a}_{bc}$) as

\begin{eqnarray}\label{connection}
{{\Gamma^a}_{bc}}^{(q)} =
{\Gamma^a}_{bc}
+ \frac{1}{2} \, q^{ad} \,
\big(-\nabla_d q_{bc} + 2 \nabla_{\left(b\right.} q_{\left.c\right)d}\big)
\end{eqnarray}
\cite{KotG},
with ${\Gamma^a}_{bc}$ evaluated at $x$.
For time or space separations,
if we combine
properties
$i)$ and $ii)$ above
with a specific requirement
concerning the Green function,
we fix $q_{ab}$.
The requirement is 
that in any maximally symmetric spacetime
the qmetric Green's function $G_{(q)}$ 
of the qmetric d'Alembertian $\Box_{(q)} = \nabla^{(q)}_a \nabla^a_{(q)}$
be given at $x$ by
$G_{(q)} = G(S)$ 
[where $G$ 
is the Green's function of $\Box$,
and $G(S)$ is $G$ evaluated at that $\tilde x$ on the geodesic
such that $\sigma^2(\tilde x, x') = S$].
This uniquely fixes $q_{ab}$ 
as a function of $g_{ab}$ and the tangent $t^a$ to the geodesic at $x$,
and of $\sigma^2$ and $S$
\cite{KotE, KotG, KotI}.
For null separations,
$q_{ab}$ is uniquely fixed by
properties
$i)'$ and $ii)'$,
joined with the requirement that, in the proximity of $x$ 
(on the null geodesic $\gamma$ from $x'$ through $x$, $G$ diverges), 
$G_{(q)} = G(\tilde\lambda)$
[where $G(\tilde\lambda)$ is $G$ 
evaluated nearby that point $\tilde x \in \gamma$ such that 
$\lambda(\tilde x, x') = \tilde\lambda$] \cite{PesN}. 

Focusing on the null case in our specific geometrical setup,
if we choose $P$ as $x'$ and $p$ as $x$,
$q_{ab}(p, P)$ reads \cite{PesN}

\begin{eqnarray}\label{tredenus2_06.11}
q_{ab} = A \, g_{ab} -\frac{1}{2} \,
\Big(\frac{1}{\alpha} - A\Big) (l_a m_b + m_a l_b).
\end{eqnarray}
Also,
from 
$q^{ab} q_{bc} = {\delta^a}_c$,

\begin{eqnarray}\label{tredenus2_06.12}
q^{ab} = \frac{1}{A} \, g^{ab} + \frac{1}{2} \,
          \Big(\frac{1}{A} - \alpha\Big) (l^a m^b + m^a l^b).
\end{eqnarray}
Here, the $g_{ab}$-null vectors $l^a$ and $m^a$ are as defined 
in the previous section
(with $l_a \equiv g_{ab} l^b$ and $m_a \equiv g_{ab} m^b$; as a general 
rule,
any index of a ordinary-metric
tensor is raised and lowered using $g^{ab}$ and $g_{ab}$
and any index of a qmetric tensor is raised and lowered using 
$q^{ab}$ and $q_{ab}$) 
then 
they are $g_{ab}$-orthogonal to $\Sigma(P, \lambda)$
with
$
g_{ab} l^a m^b = -2.
$
All vectors and tensors on the rhs of Eqs. (\ref{tredenus2_06.11}) and
(\ref{tredenus2_06.12}) are evaluated at $p$.
The quantities $\alpha = \alpha(p, P)$ and $A = A(p, P)$ 
are instead biscalars, given by the expressions

\begin{eqnarray}\label{alpha_qab}
\alpha = \frac{d\lambda}{d\tilde\lambda},
\end{eqnarray}
with the derivative taken at $p$,
and

\begin{eqnarray}\label{A_qab}
A = \frac{\tilde\lambda^2}{{\lambda}^2} \,
  \bigg(\frac{\Delta}{\tilde\Delta}\bigg)^{\frac{2}{D-2}},
\end{eqnarray}
where

\begin{eqnarray}\label{vanVleck}
\Delta(p, P) = - \frac{1}{\sqrt{g(p) g(P)}} \, 
{\rm det}\Big[-\nabla^{(p)}_a \nabla^{(P)}_b \frac{1}{2} \sigma^2(p, P)\Big]
\end{eqnarray}
is the van Vleck determinant 
\cite{vVl, Mor, DeWA, DeWB} (see Refs. \cite{Xen, VisA, PPV}),
and

\begin{eqnarray}\label{tilde_vanVleck}
\tilde\Delta(p, P) = \Delta(\tilde p, P), 
\end{eqnarray}
with
$\tilde p \in \gamma$ such that $\lambda(\tilde p, P) = \tilde\lambda$,
with both $\Delta$ and $\tilde\Delta$ being biscalars.

Equation (\ref{vanVleck}) has the quadratic interval
(doubly) differentiated along directions also off $L$.
Since the quadratic interval
has its own behavior off $L$ 
which is different in $M^\sharp$ and $M$
(and independent of our chosen extensions in $M$),
in the formula above in principle
we should have written $\sigma^{\sharp 2}$ in place of
$\sigma^2$,
thus distinguishing $M^\sharp$ and $M$.
However,
the fact that the van Vleck determinant has a direct geometrical
interpretation as the ratio of the density of geodesics
at $p$ from $P$ in the actual spacetime and the density in the would-be
flat space \cite{VisA} guarantees that
$
{\rm det}\Big[-\nabla^{(p)}_a \nabla^{(P)}_b 
\frac{1}{2} \sigma^2(p, P)\Big]^\sharp
=
{\rm det}\Big[-\nabla^{(p)}_a \nabla^{(P)}_b 
\frac{1}{2} \sigma^2(p, P)\Big],
$
for this ratio is a quantity given
with the congruence,
and the congruence is left untouched when going from $M$ to $M^\sharp$.
We have more: since $\Delta(p, P)$ is a fixed quantity
independent of the specific null congruence we might consider,
we would obtain this same ratio at $p$ with
any other congruence from $P$
containing the geodesic $\gamma$ through $P$ and $p$.
In particular, this entails that $A$ is a quantity
which is assigned with the geodesic $\gamma$,
i.e. it does not depend on what happens to $l^a$ outside $\gamma$.
This same thing can be said about $\alpha$.

$\alpha$ and $A$ are defined for every $p \in L$;
we need to consider the variation of them 
when leaving $L$. 
In Eq. (\ref{connection}) indeed, 
there are derivatives of $\alpha$ and $A$ also in directions off $L$.
Concerning this,
recall that
we introduced $\lambda$ as the distance
according to observers $V^a$ parellel transported 
along the geodesics.
%
Consequently,
we write
$\partial_a \lambda = r_a$
(with $r^a \equiv l^a - V^a$).
This provides values of $\lambda$ in a neighborhood of $p \in L$,
also at points $p' \not\in L$.
In particular,
$V^a \partial_a \lambda = V^a r_a = 0$ at $p\in L$
(amounting to $\lambda = {\rm const}$ 
momentarily at $p$ along the worldline
of the observer $V^a$),
which gives
$m^a \partial_a \lambda = -l^a \partial_a \lambda + 2 V^a \partial_a \lambda
= - l^a \partial_a \lambda$ at $p \in L$.
$\alpha$ and $A$ are regarded as functions of $\lambda$.
As such,
what was just said about $\lambda$ does apply to them as well;
specifically,
$m^a \partial_a \alpha = - l^a \partial_a \alpha$
and
$m^a \partial_a A = - l^a \partial_a A$.

In the connection (\ref{connection}), 
the terms which might render
${{\Gamma^{\sharp a}}_{bcd}}^{(q)} \ne {{{\Gamma^a}_{bcd}}}^{(q)}$
are of the type
$\partial_a \alpha$ (or $\partial_a 1/\alpha$)
and
$\partial_a A$ (or $\partial_a 1/A$).
From Eq. (\ref{133_0}) as applied to $\Phi = \alpha$ and to $\Phi = A$,
we have
$(m^a \partial_a \alpha)^\sharp = - l^a \partial_a \alpha$
and
$(m^a \partial_a A)^\sharp = - l^a \partial_a A$;
this, from what was just said in the previous paragraph,
implies
$(m^a \partial_a \alpha)^\sharp = m^a \partial_a \alpha$
and
$(m^a \partial_a A)^\sharp = m^a \partial_a A$.
Since we know that the differentiation 
of $\alpha$ and $A$ along $L$ is identical in $M^\sharp$ and $M$,
this means that
$(\partial_a \alpha)^\sharp = \partial_a \alpha$
and
$(\partial_a A)^\sharp = \partial_a A$,
and guarantees that
${{\Gamma^{\sharp a}}_{bcd}}^{(q)} = {{\Gamma^a}_{bcd}}^{(q)}$.

We now
consider what our geometric configuration looks like
in the qmetric framework.
We know, by construction,
that 
any null geodesic $\gamma$ 
with affine parameter $\lambda$ turns out to be,
according to the qmetric,
an affinely-parameterized null geodesic $\gamma_{(q)}$
with qmetric-affine parameter $\tilde\lambda$.
Its qmetric tangent vector $l^a_{(q)}$ is 
$l^a_{(q)} = \frac{dx^a}{d\tilde\lambda} 
= \frac{d\lambda}{d\tilde\lambda} \, \frac{dx^a}{d\lambda}
= \alpha \, l^a$, 
which can be readily verified to be qmetric-null,
i.e. 
$q_{ab} \, l^a_{(q)} l^b_{(q)} = 0$.
As a consequence, 
the qmetric-null hypersurface $L_{(q)}$ swept by vectors $l^a_{(q)}$,

\begin{eqnarray}
\nonumber
L_{(q)} = \big\{p \in M: 
\ S(p, P)=0 \ {\rm and} \ p \ {\rm is \ in \ the \ future \ of} 
\ P\big\},
\end{eqnarray}
does coincide with $L$, $L_{(q)} = L$,
and indeed,
at any $p$ in the future of $P$,
$\sigma^2(p, P) = 0 \Leftrightarrow S(p, P) = 0$.
$l^a_{(q)}$ is clearly qmetric-normal to $L_{(q)}$ since,
from $0 = \frac{dS}{d\tilde\lambda} 
= \frac{dx^a}{d\tilde\lambda} \, \partial_a S$,
$l_a^{(q)}$ must be parallel to the (qmetric)
null normal $\partial_a S$.
$L_{(q)}$ is locally
$L_{(q)} = \gamma_{(q)} \times \Sigma_{(q)}(P, \tilde\lambda)$,
where, for any fixed $\tilde\lambda$,

\begin{eqnarray}
\nonumber
\Sigma_{(q)}(P, \tilde\lambda) =
\big\{p \in L_{(q)}: 
\ \tilde\lambda(p, P) = \tilde\lambda \, (> 0)
\big\},
\end{eqnarray}
and can be mapped through coordinates
$(\tilde\lambda, \theta^A)$ where the same $\theta^A$ as before
label the different geodesics $\gamma_{(q)}$.
We see that
$\Sigma_{(q)}(P, \tilde\lambda) = \Sigma(P, \lambda)$,
where the affine interval $\lambda$ from $P$ is that
corresponding to the given $\tilde\lambda$
(at any $p \in L_{(q)}$,
$\tilde\lambda(p, P) = \tilde\lambda \Leftrightarrow 
\lambda(p, P) = \lambda$).

Introducing
${m'}^a_{(q)} = dx^a/d\tilde\nu$
as an auxiliary qmetric-null vector
satisfying 
${{m'}_a}^{(q)} \, {e^a_A}^{(q)} = 0$
and
${{m'}_a}^{(q)} \, l^a_{(q)} = -2$
[${e^a_A}^{(q)} \equiv 
({\partial x^a}/{\partial \theta^A})_{\tilde\lambda = {\rm const}}
=  ({\partial x^a}/{\partial \theta^A})_{\lambda = {\rm const}} 
= e^a_A$ \cite{PesN}],
it is uniquely determined
as
$ {m'}^a_{(q)} = m^a$ (with $d\tilde\nu = d\nu$), 
${m'}_a^{(q)} = q_{ab} \, {m'}^b_{(q)} = (1/\alpha) \, m_a$,
and parallel transported along each $\gamma_{(q)}$
(Appendix \ref{AppB}).
The qmetric $h_{ab}^{(q)}$ induced on the surface $\Sigma_{(q)}(P, \tilde\lambda)$
is given by

\begin{eqnarray}\label{179_1}
h_{ab}^{(q)} =
q_{ab} + \frac{1}{2} l_a^{(q)} {m'}_b^{(q)} + \frac{1}{2} {m'}_a^{(q)} l_b^{(q)},
\end{eqnarray}
cf. Ref. \cite{PesN}.

In exact analogy with the $g_{ab}$ case,
we have the qmetric curvature transverse fields

\begin{eqnarray}
\nonumber
K_{ab}^{(q)} &\equiv& {{h^c}_a}^{(q)} {{h^d}_b}^{(q)} \, \nabla_c^{(q)} l_d^{(q)} \\
\nonumber
{\bar K}_{ab}^{\prime (q)} &\equiv&  {{h^c}_a}^{(q)} {{h^d}_b}^{(q)} \,
\nabla_c^{(q)} {m'}_d^{(q)},
\end{eqnarray}
and the qmetric Riemann tensor 
${{R_{\Sigma (q)}}^{\, a}}_{bcd}$
intrinsic to 
$\Sigma_{(q)}(P, \tilde\lambda)$ defined by

\begin{eqnarray}
\nonumber
{Y^a_{(q)}}_{{\tilde \|} cd} - {Y^a_{(q)}}_{{\tilde \|} dc} 
=
- {{R_{\Sigma (q)}}^{\, a}}_{bcd} \, Y^b_{(q)},
\end{eqnarray}
$Y^a_{(q)} \in T(\Sigma_{(q)})$,
where ${\tilde \|}$
means covariant differentiation with respect to the connection
relative to $h_{ab}^{(q)}$.
From the considerations above concerning the connection,
we know that these transverse fields are the same in $M^\sharp$ and $M$.

Given these circumstances,
one would like to carry out in the qmetric spacetime
the derivation of the Ricci scalar presented in the previous section.
%
One peculiarity of $M^\sharp$ and $M$ with the qmetric,
is that the q-vectors have their own assigned behaviors off $L_(q)$ (read $L$)
which they inherit from their definitions in terms 
of the vectors $l^a$, $m^a$.
From the considerations in the paragraph below Eq. (\ref{112_4})
we know that, concerning the possibility to determine the Ricci scalar 
in a qmetric spacetime $\tilde M_{(q)}$ in which the qmetric Ricci scalar
is completely determined by $L_{(q)}$, this is not an issue,
for $l^a_{(q)}$ and ${m'}^a_{(q)}$ have still qmetric parallel transport
along the geodesics and have $l^a_{(q)} {m'}_a^{(q)} = -2$.
What is left to be seen however,
is what kind of relation links
differentiation along ${m'}^a_{(q)}$ and differentiation along $l^a_{(q)}$
as determined by their definitions,
in other words if, taking $l^a_{(q)}$ as playing the role of $l^a$ in
qmetric spacetime, ${m'}^a_{(q)}$ actually plays
the role of $m^a$.
%

Considering a scalar $\Phi$
for which $m^a \partial_a \Phi = - l^a \partial_a \Phi$ in $M$,
we get

\begin{eqnarray}\label{174_4}
{m'}^a_{(q)} \, \partial_a^{(q)} \Phi 
&=&
m^a \partial_a \Phi 
\nonumber \\
&=&
- l^a \partial_a \Phi
\nonumber \\
&=&
- \frac{d\tilde\lambda}{d\lambda} \, \frac{d\Phi}{d\tilde\lambda}
\nonumber \\
&=&
- \frac{1}{\alpha} \, l^a_{(q)} \, \partial_a^{(q)} \Phi
\end{eqnarray}
with $p \in L_{(q)}$,
where we used $\partial_a^{(q)} = \partial_a$.
%
We see that we must have
$d\Phi/d\tilde\nu = - (1/\alpha) \, d\Phi/d\tilde\lambda$,
i.e. that differentiation along ${m'}^a_{(q)}$ is
$-1/\alpha$ times differentiation along $l^a_{(q)}$.

We are now in a position
to appreciate
that the role the vector ${m'}^a_{(q)}$
plays in qmetric spacetime
does not coincide with the role played by $m^a$ in ordinary spacetime.
It is instead exactly the role 
which in the ordinary spacetime is played
by the vector $\hat m^a$ 
of Eq. (\ref{hat_133_0})
when taking $\mu = \alpha$;
this fact appears to be an expression of
the intrinsic asymmetry of the qmetric
along the radial direction, 
entailed by the requirement of
a finite limiting distance at coincidence.
Thus,
we can write the zero-point-length Gauss--Codazzi relation for
null equigeodesic hypersurfaces 
from Eq. (\ref{179_8}).
We get  

\begin{eqnarray}\label{180_4}
{\tilde R_{(q)}}
&=&
R_{\Sigma (q)} + \alpha \, K_{(q)} {\bar K'_{(q)}} + \alpha 
\, K^{ab}_{(q)} {\bar K_{ab}^{\prime (q)}}
+ \alpha \, \big({m'}_{(q)}^a \partial_a^{(q)} K_{(q)}\big)^\sharp 
+ l^a_{(q)} \partial_a^{(q)} \big(\alpha {\bar K'_{(q)}}\big)
\nonumber \\
&=&
R_{\Sigma (q)} + \alpha \, K_{(q)} {\bar K'_{(q)}} + \alpha 
\, K^{ab}_{(q)} {\bar K_{ab}^{\prime (q)}}
+ l^a_{(q)} \partial_a^{(q)} \big(\alpha {\bar K'_{(q)}} - K_{(q)}\big),
\end{eqnarray}
with $K_{(q)} = {{K_a}^a}^{(q)} \equiv q^{ab} K_{ab}^{(q)}$,
$\bar K'_{(q)} = \bar K^{\prime \, \, b {(q)}}_b$,
and $\bar K_{ab}^{\prime (q)}$ playing the role of $\hat K_{ab}$.


\section{Minimum-length Ricci scalar for null separated events}

In a qmetric spacetime $\tilde M_{(q)}$ such that 
the minimum-length Ricci scalar ${\tilde R}_{(q)}$
is completely determined by the null hypersurface $L_{(q)}$,
Eq. (\ref{180_4}) provides an expression for
$\tilde R_{(q)}$ in terms of other qmetric quantities.
We proceed now to try to gain an expression for ${\tilde R_{(q)}}$
in terms of objects which live
in ordinary spacetime, i.e. with reference to the classical metric $g_{ab}$.
%
This involves expressing the rhs of Eq. (\ref{180_4})
in terms of the quantities which define the qmetric in ordinary spacetime,
namely $\alpha$ and $A$.
Let us start from the term $R_{\Sigma (q)}$.
From
Eq. (\ref{179_1})
we get

\begin{eqnarray}
h^{(q)}_{ab} = A \ h_{ab}
\end{eqnarray}
(cf. Ref. \cite{PesN}),
i.e. $h^{(q)}_{ab}$  and $h_{ab}$ are conformally related,
in analogy to what happens 
in the timelike/spacelike case \cite{KotG}.
Since the conformal factor $A$ is constant
on $\Sigma$, this gives (cf. e.g. Ref. \cite{HawB})

\begin{eqnarray}\label{125_4}
R_{\Sigma (q)} =
\frac{1}{A} \, R_\Sigma.
\end{eqnarray}

$K^{(q)}_{ab}$ turns out to be

\begin{eqnarray}\label{127_1}
K_{ab}^{(q)} &=& 
{{h^c}_a}^{(q)} {{h^d}_b}^{(q)} \, \nabla_c^{(q)} l_d^{(q)}
\nonumber \\
&=&
K_{ab} + \frac{1}{2} \, \alpha \, {h^c}_a {h^d}_b \, l^l \nabla_l q_{cd}
-\frac{1}{2} \, \alpha \, {h^c}_a {h^d}_b \, l^l 
(\nabla_c q_{dl} + \nabla_d q_{cl})
\nonumber \\
&=&
K_{ab} + \frac{1}{2} \, \alpha \, \frac{dA}{d\lambda} \, h_{ab}
- (1-\alpha A) \, K_{ab}
\nonumber \\
&=&
\alpha A \, K_{ab} + \frac{1}{2} \, \alpha \frac{dA}{d\lambda} \, h_{ab},
\end{eqnarray}
with the derivation detailed in Appendix \ref{AppC}. 
This gives

\begin{eqnarray}\label{134_2}
K_{(q)} &=&
q^{ab} K_{ab}^{(q)}
\nonumber \\
&=&
\alpha \, K + \frac{1}{2} \, (D-2) \, \alpha \, \frac{d}{d\lambda} \ln A, 
\end{eqnarray}
where in the second step we used $l^a K_{ab} = 0$, $l^a h_{ab} = 0$.

As for ${\bar K}_{ab}^{\prime (q)}$, we get
(the derivation is
spelled out in Appendix \ref{AppC}) 

\begin{eqnarray}\label{132_3}
{\bar K}_{ab}^{\prime (q)} &=&  
{{h^c}_a}^{(q)} {{h^d}_b}^{(q)} \, \nabla_c^{(q)} {m'}_d^{(q)}
\nonumber \\
&=&
A \, {\bar K}_{ab} - \frac{1}{2} \, \frac{dA}{d\lambda} \, h_{ab}.
\end{eqnarray}
This gives

\begin{eqnarray}\label{135_1}
{\bar K'}_{(q)} &=&
q^{ab} {\bar K}_{ab}^{\prime (q)}
\nonumber \\
&=&
{\bar K} - \frac{1}{2} \, (D-2) \, \frac{d}{d\lambda} \ln A, 
\end{eqnarray}
where we used $l^a {\bar K}_{ab} = 0$.

Using Eqs. (\ref{127_1})--(\ref{135_1}), 
for various terms in Eq. (\ref{180_4}) we get 

\begin{eqnarray}
\nonumber
- l^a_{(q)} \partial_a^{(q)} K_{(q)} &=&
- \alpha \, \frac{d\alpha}{d\lambda} \, K
- \alpha^2 \, \frac{dK}{d\lambda}
- \frac{1}{2} (D-2) \, \alpha \, \frac{d\alpha}{d\lambda} \, 
\frac{d}{d\lambda} \ln A
-\frac{1}{2} (D-2) \, \alpha^2 \, \frac{d^2}{d\lambda^2} \ln A, 
\nonumber \\
l^a_{(q)} \partial_a^{(q)} \big(\alpha {\bar K'_{(q)}}\big) &=&
\alpha \, l^a_{(q)} \nabla_a^{(q)} {\bar K'_{(q)}}
+ \big(l^a_{(q)} \nabla_a^{(q)} \alpha\big) \, {\bar K'_{(q)}}
\nonumber \\
&=&
\alpha^2 \, \frac{d\bar K}{d\lambda} 
- \frac{1}{2} (D-2) \, \alpha^2 \, \frac{d^2}{d\lambda^2} \, \ln A 
+ \alpha \, \frac{d\alpha}{d\lambda} \,
\Big({\bar K} - \frac{1}{2} (D-2) \frac{d}{d\lambda} \ln A\Big),
\nonumber \\
\label{K_barK}
\alpha \, K^{ab}_{(q)} {\bar K_{ab}^{\prime (q)}} &=&
\alpha^2 \, K^{ab} {\bar K_{ab}}
+ \frac{1}{2} \, \alpha^2 \Big(\frac{d}{d\lambda} \ln A\Big) 
({\bar K} - K)
- \frac{1}{4} (D-2) \, \alpha^2 \Big(\frac{d}{d\lambda} \ln A\Big)^2
\end{eqnarray}
(the calculation of the last equality is detailed in Appendix \ref{AppC}),
and Eq. (\ref{180_4}) becomes

\begin{eqnarray}\label{182_3}
{\tilde R_{(q)}} &=& 
\frac{1}{A} \, R_\Sigma
+ \alpha \, \frac{d\alpha}{d\lambda} \, ({\bar K} - K)
+ \alpha^2 \, \Big(\frac{d\bar K}{d\lambda} - \frac{dK}{d\lambda}\Big)
- (D-2) \, \alpha \, \frac{d\alpha}{d\lambda} \, 
\frac{d}{d\lambda} \ln A
\nonumber \\
&-& (D-2) \, \alpha^2 \, \frac{d^2}{d\lambda^2} \ln A
+ \alpha^2 \, K {\bar K} + \frac{1}{2} (D-1) \, \alpha^2 \,
\Big(\frac{d}{d\lambda} \ln A\Big) ({\bar K} - K)
\nonumber \\
&-& \frac{1}{4} (D-2) (D-1) \, \alpha^2 \, 
\Big(\frac{d}{d\lambda} \ln A\Big)^2
+ \alpha^2 \, K^{ab} {\bar K_{ab}}. 
\end{eqnarray}



\section{Coincidence and $L_0 \to 0$ limit}

Having the expression (\ref{182_3}) for $\tilde R_{(q)}$ for the null case,
namely the expression for the qmetric Ricci scalar
near a point $P$ in the case where $\tilde R_{(q)}$ is completely determined by
the null geodesics emerging from $P$,
we can ask what this expression becomes in the coincidence limit
$p \to P$.
%
One might naively expect that it turns out to be the Ricci scalar $R$ at $P$
plus a certain quantity $\eta$, dependent on $L_0$, 
with $\eta$ vanishing in the $L_0 \to 0$ limit.  
In the case of the qmetric Ricci scalar for time- or space-separated events 
there already exists the remarkable result
that this is not the case. 
What has been found is   
$\lim_{L_0 \to 0} \, (\lim_{p \to P} R_{(q)}) = \varepsilon D \, R_{ab} t^a t^b
\not= R$,
where $t^a$ is the normalized tangent (at $P$)
to the geodesics connecting $P$ and $p$ \cite{Pad01, Pad02, KotI}.
Our aim here is to see whether something similar
happens in the null case, 
thing which, in view of these results, 
might somehow
be by now expected
(by analogy with the spacelike/timelike case
or from entropy density considerations \cite{Pad01, Pad02}).

We begin by noting that, from Eq. (\ref{112_4bis}), 
$R_\Sigma$ can be expressed entirely
in terms of the other quantities already present in Eq. (\ref{182_3}) 
and of $R^\sharp$ as 

\begin{eqnarray}\label{163_2}
R_\Sigma &=&
R^\sharp
- K {\bar K} - K^{ab} {\bar K}_{ab}
- l^a \partial_a ({\bar K} - K)
\nonumber \\
&=&
R^\sharp + K^2 + K^{ab} K_{ab} + 2 \, \frac{dK}{d\lambda}
- \Big(K \epsilon + K^{ab} \epsilon_{ab} + \frac{d\epsilon}{d\lambda}\Big)
\nonumber \\ 
&=&
R^\sharp + K^2 - K^{ab} K_{ab} - 2 R_{ab} l^a l^b 
- \Big(K \epsilon + K^{ab} \epsilon_{ab} + \frac{d\epsilon}{d\lambda}\Big),
\end{eqnarray}
where we introduced
$\epsilon_{ab} \equiv K_{ab} + {\bar K_{ab}}$
with $\epsilon \equiv g^{ab} \epsilon_{ab} = h^{ab} \epsilon_{ab}$.
In the third equality, we used (see Appendix \ref{AppD})
\begin{eqnarray}\label{162_1}
\frac{dK}{d\lambda} =
- K^{ab} K_{ab} - {R_{ab}^\sharp} \, l^a l^b.
\end{eqnarray}
Here, 
consideration of Raychaudhuri's equation
(which gives the rate of expansion of the congruence
in terms of the parameters characterizing the congruence
and of only the quantity $R_{ab} l^a l^b$)
assures that

\begin{eqnarray}\label{Riccill}
R_{ab}^\sharp \, l^a l^b =
R_{ab} \, l^a l^b,
\end{eqnarray}
for the congruence is left untouched when going 
from $M$ to $M^\sharp$. 

Since $A$ is expressed in terms of the van Vleck determinant biscalar,
we see that what we need in Eq. (\ref{182_3})
is knowledge of the behavior
of this, as well as of $K_{ab}$ and of the terms in parentheses
in Eq. (\ref{163_2}),
for $p$ near $P$,
and to deal with the $R^\sharp$ term.

As for the van Vleck determinant,
from Ref. \cite{Xen} we get

\begin{eqnarray}\label{164_1}
\Delta^{1/2}(p, P) =
1 + \frac{1}{12} \, \lambda^2 R_{ab} \, l^a l^b + {\cal O}(\lambda^3),
\end{eqnarray}
with $R_{ab}$ and the vector $l^a$ evaluated at $p$
[incidentally, we verify here that the $\lambda^2$ term in the expansion
is the same in $M^\sharp$ and $M$, consistently
with the comments in the paragraph below Eq. (\ref{tilde_vanVleck})].
This also gives

\begin{eqnarray}\label{165_1}
\tilde\Delta^{1/2}(p, P) =
1 + \frac{1}{12} \, 
\tilde\lambda^2 \big(R_{ab} \, l^a l^b\big)_{|\tilde p} 
+ {\cal O}(\tilde\lambda^3).
\end{eqnarray}

As for $K_{ab}$, 
we have

\begin{eqnarray}\label{142_0}
K_{ab}(p) &=& 
{h^c}_a \, {h^d}_b \, \nabla_c l_d
\nonumber \\
&=&
{h^c}_a \, {h^d}_b \, 
\nabla_c\bigg(\frac{1}{\lambda} \nabla_d\Big(\frac{\sigma^2}{2}\Big)\bigg)
\nonumber \\
&=&
{h^c}_a \, {h^d}_b \,
\bigg[\Big(\nabla_c \frac{1}{\lambda}\Big) \nabla_d\Big(\frac{\sigma^2}{2}\Big)
+ \frac{1}{\lambda} \, \nabla_c \nabla_d\Big(\frac{\sigma^2}{2}\Big)\bigg]
\nonumber \\
&=&
\frac{1}{\lambda} \, {h^c}_a \, {h^d}_b \, 
\nabla_c \nabla_d\Big(\frac{\sigma^2}{2}\Big).
\end{eqnarray}
The second equality here,
comes about since,
for $p'$ near $p$, but not exactly on the null geodesic 
$\gamma$ through $P$ and $p$,
$\nabla_d\Big({\sigma^2(p', P)}/{2}\Big)$
can be usefully expressed as
$\nabla_d\Big({\sigma^2(p', P)}/{2}\Big) 
= \lambda \, l_d + \nu \, m_d$
(with $\lambda$ and $\nu$ meant 
as curvilinear null coordinates of $p'$, $\nu = 0$ on $\gamma$) \cite{VisA}.
The vector fields 
$(1/\lambda) \, \nabla_d({\sigma^2}/{2})$
and
$l_d$
do coincide then
at any $p \in L$ but behave differently
when going off $L$.
The projections of their derivatives on $\Sigma(P, \lambda)$ however,
i.e. the quantities 
${h^c}_a {h^d}_b \nabla_c \big[\frac{1}{\lambda} 
\, \nabla_d({\sigma^2}/{2})\big]$
and
${h^c}_a {h^d}_b \nabla_c l_d$
do coincide at any $p \in L$,
for we know they do not depend on which is the extension
of the vector fields off $L$.
The fourth equality in Eq. (\ref{142_0}) is a consequence of
$\nabla_c \frac{1}{\lambda} \to \big(\nabla_c \frac{1}{\lambda}\big)_{|\Sigma}$ 
for $\nu \to 0$ 
and of ${h^c}_a \nabla_c ({1}/{\lambda})$ being the gradient 
of $1/\lambda$ on $\Sigma$, where $\lambda ={ \rm const}$.
We have
$\big[\nabla_c \big(\frac{1}{\lambda} 
\nabla_d(\sigma^2/2)\big)\big]^\sharp
\ne
\big[\nabla_c \big(\frac{1}{\lambda} 
\nabla_d(\sigma^2/2)\big)\big],
$
as well as
$\big[\nabla_c \nabla_d (\sigma^2/2)\big]^\sharp 
\ne
\nabla_c \nabla_d (\sigma^2/2);
$
in the equalities above we have not inserted however
any $^\sharp$ symbol because, after the projection, 
the quantities we obtain turn out
to be the same in $M^\sharp$ and $M$:
${h^c}_a {h^d}_b \big[\nabla_c \big(\frac{1}{\lambda} 
\nabla_d(\sigma^2/2)\big)\big]^\sharp
=
{h^c}_a {h^d}_b \nabla_c l_d
=
{h^c}_a {h^d}_b \big[\nabla_c \big(\frac{1}{\lambda} 
\nabla_d(\sigma^2/2)\big)\big]
$
and
${h^c}_a {h^d}_b \big[\nabla_c \nabla_d (\sigma^2/2)\big]^\sharp
=
{h^c}_a {h^d}_b \nabla_c l_d
=
{h^c}_a {h^d}_b \nabla_c \nabla_d (\sigma^2/2).
$

Equation (\ref{142_0}) shows that what we need is the expansion 
for $\nabla_c \nabla_d({\sigma^2}/{2})$ for $p$ near $P$.
From Ref. \cite{Xen},
we get

\begin{eqnarray}\label{141_3}
\nabla_a \nabla_b\Big(\frac{\sigma^2}{2}\Big) =
g_{ab} - \frac{1}{3} \, \lambda^2 \, E_{ab}
+ {\cal O}(\lambda^3),
\end{eqnarray}
where 
$E_{ab} \equiv R_{ambn} \, l^m l^n$
and the quantities on the rhs are evaluated at $p$.
As for $K_{ab}$, 
this gives

\begin{eqnarray}\label{142_3}
K_{ab}(p) =
\frac{1}{\lambda} \, h_{ab}
- \frac{1}{3} \, \lambda \,
{h^c}_a \, {h^d}_b \, E_{cd}
+ {\cal O}(\lambda^2),
\end{eqnarray}
an expression quite similar to (the leading terms of) 
the expansion found
for the extrinsic curvature of equigeodesic surfaces
for timelike/spacelike cases \cite{KotG, KotI}.
This also gives
\begin{eqnarray}
K 
&=&
(D-2) \, \frac{1}{\lambda} 
- \frac{1}{3} \, \lambda E
+ {\cal O}(\lambda^2).
\end{eqnarray}
%
In Eq. (\ref{142_3}), one might want to check
that the second term of the rhs turns out to be the same
in $M^\sharp$  and $M$,
i.e. that

\begin{eqnarray}\label{Q1_70_4} 
{h^c}_a \, {h^d}_b \, E_{cd}^\sharp
= {h^c}_a \, {h^d}_b \, E_{cd}.
\end{eqnarray} 
This is done in Appendix \ref{AppE}. 

As for the terms in parentheses in Eq. (\ref{163_2}),
it can be verified that

\begin{eqnarray}\label{epsilon}
\epsilon_{ab} \to 0,
\ \ \ \frac{d\epsilon}{d\lambda} \to 0,
\end{eqnarray}
for $\lambda \to 0$
(this is detailed in Appendix \ref{AppF}),
and then $(\epsilon/\lambda) \to 0$.
We get thus

\begin{eqnarray}
K \epsilon = 
(D-2) \, \frac{\epsilon}{\lambda} + \epsilon \, {\cal O}(\lambda)
\, \to 0
\nonumber
\end{eqnarray}
and

\begin{eqnarray}
K^{ab} \epsilon_{ab} 
=
\frac{1}{\lambda} \, h^{ab} \epsilon_{ab} + E^{ab} \epsilon_{ab} 
\, {\cal O}(\lambda) 
=
\frac{\epsilon}{\lambda} + E^{ab} \epsilon_{ab} 
\, {\cal O}(\lambda)
\, \to 0 
\end{eqnarray}
for $\lambda \to 0$.
This shows that the terms in parentheses have no effect
in the expression of Eq. (\ref{182_3}) in the same limit,
i.e.
that

\begin{eqnarray}\label{183_3}
\lim_{\lambda \to 0} {\tilde R_{(q)}} 
&=&
\lim_{\lambda \to 0}
\Bigg\{\frac{1}{A} \, R_\Sigma
- 2 \, \alpha \, \frac{d\alpha}{d\lambda} \, K
+ 2 \, \alpha^2 \, R_{ab} \, l^a l^b
- (D-2) \, \alpha \, \frac{d\alpha}{d\lambda} \, 
\frac{d}{d\lambda} \ln A
- (D-2) \, \alpha^2 \, \frac{d^2}{d\lambda^2} \ln A
\nonumber \\
&-& \frac{1}{4} (D-2) (D-1) \, \alpha^2 \, 
\Big(\frac{d}{d\lambda} \ln A\Big)^2
- \alpha^2 K^2
+ \alpha^2 K^{ab} K_{ab}
- (D-1) \, \alpha^2 \,
\Big(\frac{d}{d\lambda} \ln A\Big) K\Bigg\}.
\end{eqnarray}

Using Eq. (\ref{A_qab}) for $A$, 
the expression (\ref{163_2}) for $R_\Sigma$ and the expansions
(\ref{164_1}), (\ref{165_1}) and (\ref{142_3}), 
the first term in the expression on the rhs of Eq. (\ref{183_3}) is found to be

\begin{eqnarray}\label{166_3}
\frac{1}{A} \, R_\Sigma &=&
(D-2) (D-3) \,
\frac{1}{\tilde\lambda^2} \,
\tilde\Delta^{\frac{2}{D-2}} 
+ {\cal O}(\lambda^2)
\nonumber \\
&=&
(D-2) (D-3) \, \frac{1}{\tilde\lambda^2} \,
+ \frac{D-3}{3} \, E(\tilde p)
+ {\cal O}(\tilde\lambda, \lambda^2),
\end{eqnarray}
with
$E = {E^a}_a = R_{ab} \, l^a l^b$,
and where $R^\sharp$ 
enters the ${\cal O}(\lambda^2)$ term
and gives then no contribution
in the $\lambda\to 0$ limit.
%
From Eq. (\ref{142_3}) we also get

\begin{eqnarray}\label{144}
K^2
&=&
(D-2)^2 \, 
\frac{1}{\lambda^2}
- \frac{2}{3} (D-2) \, E 
+ {\cal O}(\lambda),
\\
K^{ab} K_{ab}
&=&
(D-2) \, \frac{1}{\lambda^2}
- \frac{2}{3} \, E
+ {\cal O}(\lambda).
\end{eqnarray}

As for the terms in Eq. (\ref{183_3}) containing derivatives
of $\ln A$, 
we can use the following expansions 
[from Eqs. (\ref{164_1}) and (\ref{165_1})]: 

\begin{eqnarray}\label{167_3}
\frac{d}{d\lambda} \ln A 
&=&
\frac{2}{\tilde\lambda} \, \frac{1}{\alpha} - \frac{2}{\lambda}
- \frac{2}{3 (D-2)} \, \tilde\lambda \, \frac{1}{\alpha} \, E(\tilde p)
+ \frac{2}{3 (D-2)} \, \lambda \, E(p)
+{\cal O}(\tilde\lambda^2, \lambda^2)
\\
\frac{d^2}{d\lambda^2} \ln A
&=&
- \frac{2}{\tilde\lambda^2} \, \frac{1}{\alpha^2}
- \frac{2}{\tilde\lambda} \, \frac{1}{\alpha^2} \, \frac{d\alpha}{d\lambda}
+ \frac{2}{\lambda^2}
- \frac{2}{3 (D-2)} \, \frac{1}{\alpha^2} \, E(\tilde p)
+ \frac{2}{3 (D-2)} \, E(p)
+ {\cal O}(\tilde\lambda, \lambda).
\end{eqnarray}
In particular, we get

\begin{eqnarray}\label{186}
&-& \frac{1}{4} (D-2) (D-1) \, \alpha^2 \, 
\Big(\frac{d}{d\lambda} \ln A\Big)^2
- (D-1) \, \alpha^2 \,
\Big(\frac{d}{d\lambda} \ln A\Big) K
\nonumber \\
&=&
(D-2) (D-1) \, \frac{1}{\lambda^2} \, \alpha^2
- (D-2) (D-1) \, \frac{1}{\tilde\lambda^2}
+ \frac{2}{3} (D-1) \, E(\tilde p)
- \frac{2}{3} (D-1) \, \alpha^2 \, E(p)
+ {\cal O}(\tilde\lambda, \lambda).
\end{eqnarray}

Substituting Eqs. (\ref{166_3})-(\ref{186})
into the rhs of Eq. (\ref{183_3}),
for small $\lambda$ we get

\begin{eqnarray}
{\tilde R_{(q)}} &=&
\bigg[\frac{D-3}{3} + \frac{2}{3} + \frac{2}{3}(D-1)\bigg] \, E(\tilde p)
+ {\cal O}(\tilde\lambda, \lambda)
\nonumber \\
&=&
(D-1) \, \big(R_{ab} \, l^a l^b\big)_{|\tilde p} 
+ {\cal O}(\tilde\lambda, \lambda),
\end{eqnarray}
%
which gives

\begin{eqnarray}\label{coincidence_and_0}
\Big(\lim_{L_0 \to 0} \, \lim_{\lambda \to 0} {\tilde R_{(q)}}\Big)_{|P} 
&=&
(D-1) \, \big(R_{ab} \, l^a l^b\big)_{|P} \, \, .
\end{eqnarray}

This is the limiting expression we get
for the qmetric Ricci scalar at a point $P$ of a spacetime $M$
in terms of null geodesics through $P$.
It is valid any time this scalar results fully characterizable in the qmetric
by the consideration of the collection 
of all null geodesics from $P$.


\section{Comments and conclusion}

The expression on the rhs turns out to be proportional
to (the opposite of) 
a quantity that has been interpreted as
the gravitational heat density $H$ of a null surface
with normal $l^a$ \cite{PadZ, Pad20}

\begin{eqnarray}\label{paddy_heat}
H = - \frac{1}{L_{Pl}^2} \, R_{ab} \, l^a l^b ,
\end{eqnarray}
where $L_{Pl}$ is the Planck length
(units, here and below, 
make the vacuum speed of light 
$c = 1$ 
and
the reduced Planck's constant  
$\hbar = 1$,
with Einstein's equation in the form 
$G_{ab} = 8 \pi L_{Pl}^2 \, T_{ab}$,
where $G_{ab}$ is the Einstein tensor and 
$T_{ab}$ the matter energy-momentum tensor).
Equally,
Eq. (\ref{coincidence_and_0}) turns out to enter
the entropy functional $S$
mentioned above in the statistical derivation of Einstein's equation,
in the term which accounts
for the gravitational degrees of freedom,
which has the form

\begin{eqnarray}\label{paddy_dofs}
\ln \rho_{g} = 
\frac{1}{4} \, \bigg(1 - \frac{L_{Pl}^2}{2\pi} \, R_{ab} \, l^a l^b\bigg),
\end{eqnarray}
where $\rho_g$ is the density of
quantum states of spacetime at $P$ \cite{Pad20}. 

Equation (\ref{paddy_heat}) talks about a heat density 
one has to assign to a horizon, 
namely about some quantum degrees of freedom
for spacetime.
If we allow for $L_{Pl} \to 0$ in that equation,
the apparent divergence of $H$ 
is canceled by the scaling
of $R_{ab} \, l^a l^b$ itself as $L_{Pl}^2$
(from Einstein's equation).
This leads the spacetime heat content $H$, yet a notion of quantum origin, 
to be insensitive
to the actual value of $L_{Pl}$.
Writing, from Eq. (\ref{paddy_heat}), 
$H$ in terms of Newton's constant as  

\begin{eqnarray}
H = 
- \frac{1}{G} \, R_{ab} \, l^a l^b,
\end{eqnarray}
we see thus that the quantity on the right,
even if written entirely in terms of (measurable) quantities
with apparently nothing in them talking about quantum mechanics,
can be endowed
indeed with a quantum-mechanical intrinsic significance,
which we are led to think of as sort of a relic 
of a quantumness of spacetime (for it survives to $L_{Pl} \to 0$).
This corresponds 
to the view that gravity,
starting from Newton's law itself, 
ought to be considered as intrinsically
quantum \cite{PadB_2, PadB_3}. 

This is also what
the analysis in the present paper,
in analogy of what was already pointed out in Refs. \cite{Pad01, Pad02},
actually seems to suggest.
We have indeed
that,
in the limit of actual quantumness of spacetime
going to be unnoticeable,
the quantum Ricci scalar $R_{(q)}$,
as expressed through null geodesics at $P$,
becomes
$R_{ab} \, l^a l^b$,
not $R$.
The ``classical'' quantity $R_{ab} \, l^a l^b$,
center stage in horizon thermodynamics,
is then shown to have in it
a quantum relic of the Ricci scalar of the quantum spacetime.
Because the Ricci scalar 
is nothing but the Einstein-Hilbert Lagrangian,
it in turn gives further support to the view \cite{Pad20}
that gravitational dynamics 
might be better captured in terms of $R_{ab} \, l^a l^b$
than $R$, 
i.e. somehow better in terms of
thermodynamics than geometry.

{\it Acknowledgments.}
I thank Francesco Anselmo and Sumanta Chakraborty
for comments, suggestions and improvements on the draft,
as well as for providing some references. 
I thank Dawood Kothawala for having raised the point 
which this study tries to answer.

\appendix
\section{Derivation of Eqs. (\ref{107_4}) and (\ref{111_0})}\label{AppA}

We consider first the derivation of
Eq. (\ref{107_4}).
We have [working entirely in $M^\sharp$ (see text)] 

\begin{eqnarray}\label{107_1}
R_{efgh}^\sharp \, l^e m^f m^g l^h
&=&
- R_{fegh}^\sharp \, l^e m^f m^g l^h
\nonumber \\
&=&
- {R^{\sharp f}}_{egh} \, l^e m_f m^g l^h
\nonumber \\
&=&
\big(\nabla_h \nabla_g \, l^f - \nabla_g \nabla_h \, l^f\big)^\sharp \, m_f m^g l^h
\nonumber \\
&=&
\big(l^h \nabla_h \nabla_g \, l^f\big) \, m_f m^g
- \big(m^g \nabla_g \nabla_h \, l^f\big)^\sharp \, m_f l^h,
\end{eqnarray}
from
$(\nabla_h \nabla_g \, l^f - \nabla_g \nabla_h \, l^f)^\sharp =
- {R^{\sharp f}}_{egh} \, l^e$.
Now,

\begin{eqnarray}
\big(l^h \nabla_h \nabla_g \, l^f\big) \,m^g
&=&
l^h \nabla_h \big(m^g \, \nabla_g \, l^f\big)
- \big(l^h \nabla_h m^g\big) \, \nabla_g \, l^f
\nonumber \\
&=&
 l^h \nabla_h \big(m^g \, \nabla_g \, l^f\big)
\nonumber \\
&=&
0,
\end{eqnarray}
where the second equality comes from $m^a$ being parallel 
transported along $\gamma$, 
and
the third from
the vector $v^f = m^g \, \nabla_g \, l^f$ 
identically vanishing along $\gamma$
from Eq. (\ref{102_6}) as applied
to $z^f = l^f$.

In the second term of the rhs of Eq. (\ref{107_1}),
we have

\begin{eqnarray}
\big(m^g \nabla_g \nabla_h \, l^f\big)^\sharp \, l^h
&=&
\big[m^g \nabla_g \big(l^h \, \nabla_h \, l^f\big)\big]^\sharp
- \big(m^g \nabla_g l^h\big) \, \nabla_h \, l^f
\nonumber \\
&=&
\big[m^g \nabla_g \big(l^h \, \nabla_h \, l^f\big)\big]^\sharp
\nonumber \\
&=&
0,
\end{eqnarray}
where the second equality comes from
Eq. (\ref{102_6}) as applied
to $z^h = l^h$,
and the third from Eq. (\ref{102_6})
as applied to the vector $z^f = l^h \, \nabla_h \, l^f$
which is identically vanishing along $\gamma$.
We thus get Eq. (\ref{107_4}).

As concerns Eq. (\ref{111_0}),
we have

\begin{eqnarray}\label{108_2}
2 R_{ab}^\sharp \, l^a m^b
&=& 
R_{ab}^\sharp \, l^a m^b + R_{ab}^\sharp \, m^a l^b
\nonumber \\
&=&
- (m^b \nabla_b \nabla_a \, l^a)^\sharp + (m^b \nabla_a \nabla_b \, l^a)^\sharp
- l^b \nabla_b \nabla_a \, m^a + (l^b \nabla_a \nabla_b \, m^a)^\sharp
\nonumber \\
&=&
- (m^b \nabla_b \nabla_a \, l^a)^\sharp + [\nabla_a (m^b \nabla_b \, l^a)]^\sharp
- l^b \nabla_b \nabla_a \, m^a + [\nabla_a (l^b \nabla_b \, m^a)]^\sharp
-2 (\nabla_b \, l^a) \, \nabla_a \, m^b,
\end{eqnarray}
where the second equality is from
($\nabla_b \nabla_a \, v^a - \nabla_a \nabla_b \, v^a)^\sharp
= - R_{ab}^\sharp \, v^a$ for any vector $v^a$.
In Eq. (\ref{108_2}),

\begin{eqnarray}\label{108_3}
\nabla_a \, l^a 
&=&
g^{ab} {\delta^c}_a \nabla_c \, l_b
\nonumber \\
&=&
h^{ab} {\delta^c}_a \nabla_c \, l_b
- \frac{1}{2} \, l^a m^b \nabla_a \, l_b
- \frac{1}{2} \, m^a l^b \nabla_a \, l_b
\nonumber \\
&=&
h^{ab} {\delta^c}_a \nabla_c \, l_b
\nonumber \\
&=&
h^{ab} \big({h^c}_a 
- \frac{1}{2} \, l^c m_a - \frac{1}{2} \, m^c l_a\big) \,
\nabla_c \, l_b
\nonumber \\
&=&
h^{ab} {h^c}_a \nabla_c \, l_b
\nonumber \\
&=&
{K^a}_a
\nonumber \\
&=&
K,
\end{eqnarray}
where in the second and fourth equalities
we use Eq. (\ref{62_1}),
the third equality comes from $l_b$ being parallel transported
along $\gamma$ and from Eq. (\ref{102_6}) as applied to $z_b = l_b$ 
[or simply from Eqs. (\ref{100_3}) and (\ref{100_6})],
the fifth from Eq. (\ref{102_6}) (or from $h^{ab} \, m_a = 0 = h^{ab} \, l_a$),
and the penultimate equality from the definition of $K_{ab}$.
In an analogous manner, we get

\begin{eqnarray}\label{109_1}
\nabla_a \, m^a
= {\bar K}.
\end{eqnarray}

As for the last term on the rhs of (\ref{108_2}),
we have

\begin{eqnarray}\label{110_2}
\nabla_b l^a \ \nabla_a m^b
&=&
g^{af} g^{bg} \, \nabla_b l_f \ \nabla_a m_g
\nonumber \\
&=&
\big(h^{af} - \frac{1}{2} \, l^a m^f - \frac{1}{2} \, m^a l^f\big)
\big(h^{bg} - \frac{1}{2} \, l^b m^g - \frac{1}{2} \, m^b l^g\big)
\, \nabla_b l_f \ \nabla_a m_g
\nonumber \\
&=&
h^{af} h^{bg} \, \nabla_b l_f \ \nabla_a m_g
\nonumber \\
&=&
h^{af} h^{bg} \, \nabla_f l_b \ \nabla_a m_g
\nonumber \\
&=&
K_{ab} {\bar K^{ab}}.
\end{eqnarray}
Here,
the third equality stems from parallel transporting along $\gamma$
and from repeated use of Eq. (\ref{102_6}) 
[or of Eqs. (\ref{100_3}) and (\ref{100_6})];
the fourth equality comes from the relation

\begin{eqnarray}\label{72_2}
\nabla_b \, l_f
=
\nabla_f \, l_b 
+ \big(\nabla_c \, l_b -  \nabla_b \, l_c\big) \,
\frac{1}{2} \, m^c l_f
+ \big(\nabla_f \, l_c -  \nabla_c \, l_f\big) \,
\frac{1}{2} \, m^c l_b
\end{eqnarray}
(cf. e.g. Sec. 2.4.3 of Ref. \cite{PoiA}),
combined with the orthogonality of $l^a$ to $T(\Sigma)$, 
i.e., $h^{ab} \, l_a = 0$;
and the fifth equality comes from 

\begin{eqnarray}\label{71_5}
K_{ab} {\bar K^{ab}}
&\equiv&
{h^c}_a \, {h^d}_b \, h^{fa} \, h^{gb} \, \nabla_c l_d \ \nabla_f m_g
\nonumber \\
&=&
h^{fc} \, h^{gd} \, \nabla_c l_d \ \nabla_f m_g,
\end{eqnarray}
(from ${h^c}_a h^{fa} = h^{fc}$).

Substituting Eqs. (\ref{108_3})--(\ref{110_2})
into Eq. (\ref{108_2}), we get Eq. (\ref{111_0}).

\section{Calculation of ${m'}^a_{(q)}$ and proof
of its qmetric parallel transport along any $\gamma_{(q)}$}\label{AppB}


The expression ${m'}^a_{(q)} = m^a$
[right above Eq. (\ref{179_1})]
comes, uniquely, from the requirements that
${m'}^a_{(q)}$ be null and that
${{m'}_a}^{(q)} \, {e^a_A}^{(q)} = 0$
and
${{m'}_a}^{(q)} \, l^a_{(q)} = -2$.
This happens because
${m'}^a_{(q)}$ qmetric-orthogonal to  ${e^a_A}^{(q)}$ means
that ${m'}^a_{(q)}$ cannot develop components along ${e^a_A}^{(q)}$
(which is $e^a_A$) and then we must have 
${m'}^a_{(q)} = k \, m^a + c \, l^a$ with $k, c$ scalars.
The fact that
${m'}^a_{(q)}$ is qmetric-null ,
implies further that

\begin{eqnarray}
0 &=&
q_{ab} \, {m'}^a_{(q)} {m'}^b_{(q)}
\nonumber \\
&=&
\bigg[A \, g_ab - \frac{1}{2} \Big(\frac{1}{\alpha} - A\Big) 
(l_a\, m_b + m_a \, l_b)\bigg]
(k \, m^a + c \, l^a) (k \, m^b + c \, l^b)
\nonumber \\
&=&
- 4 \, k c \, \Big(3 A - \frac{2}{\alpha}\Big), 
\end{eqnarray}
which gives
$k = 0$ or $c = 0$.
From
\begin{eqnarray}
q_{ab} \, {m'}^a_{(q)} \, l^b_{(q)} =
- 2 \, k,
\end{eqnarray}
the requirement that
$q_{ab} \, {m'}^a_{(q)} \, l^b_{(q)} = -2$
implies
$k = 1$ and then $c = 0$,
which gives
${m'}^a_{(q)} = m^a$.

As for the transport along $\gamma_{(q)}$,
we get

\begin{eqnarray}
l^b_{(q)} \nabla^{(q)}_b {m'}^{(q)}_c &=&
\alpha \, l^b 
\Big[\partial_b \frac{m_c}{\alpha} 
- {{\Gamma^a}_{bc}}^{(q)} \, \, \frac{m_a}{\alpha}\Big]
\nonumber \\
&=&
l^b \nabla_b m_c 
+ \alpha \, m_c \, \frac{d}{d\lambda}\Big(\frac{1}{\alpha}\Big)
- \frac{1}{2} \, q^{ad} \, 
\big(-\nabla_d q_{bc} + 2 \nabla_{\left(b\right.} q_{\left.c\right)d}\big) \, l^b m_a
\nonumber \\
&=&
\alpha \, m_c \, \frac{d}{d\lambda}\Big(\frac{1}{\alpha}\Big)
- \frac{1}{2} \, \alpha \, m^d l^b \, 
\big(-\nabla_d q_{bc} + 2 \nabla_{\left(b\right.} q_{\left.c\right)d}\big)
\nonumber \\
&=&
\alpha \, 
\Big({\delta^b}_c + \frac{1}{2} \, l^b m_c + \frac{1}{2} \, m^b l_c\Big) \,
\nabla_b \frac{1}{\alpha}
- \frac{1}{4} \, (1 - \alpha A) \, l^b m^d \nabla_d \, (m_b l_c)
\nonumber \\
&=&
\alpha \, {h^b}_c \nabla_b \frac{1}{\alpha}
\nonumber \\
&=&
0,
\end{eqnarray}
showing that 
the transport of ${m'}^a_{(q)}$ along $\gamma_{(q)}$ is qmetric-parallel.
Here
the third step comes from 
being $l^b \nabla_b m_c = 0$
and $q^{ad} \, m_a = \alpha \, m^d$.
In the fourth step we used Eq. (\ref{100_Phi}) putting 
$\Phi = l^a l_a$ and $\Phi = m^a l_a$.
The fifth is from Eq. (\ref{102_6}) with 
$z^b = m^b$ and $z^b = l^b$ alternately.
The last step comes from
that we are taking a gradient on $\Sigma(P, \lambda)$,
an $\alpha = {\rm const}$ there.

\section{Calculation of $K_{ab}^{(q)}$, 
${\bar K}_{ab}^{\prime (q)}$, and
$K^{ab}_{(q)} {\bar K}_{ab}^{\prime (q)}$}\label{AppC}


We begin with
the expression for $K_{ab}^{(q)}$
[Eq. (\ref{127_1}) in the main text]. 
Our starting point is
\begin{eqnarray}
K_{ab}^{(q)} &=& 
{{h^c}_a}^{(q)} {{h^d}_b}^{(q)} \, \nabla_c^{(q)} l_d^{(q)}.
\nonumber
\end{eqnarray}
Here,

\begin{eqnarray}\label{81_8}
{{h^d}_b}^{(q)}
&=&
q^{da} \, h_{ab}^{(q)}
\nonumber \\
&=&
\bigg[\frac{1}{A} \, g^{da} 
+ \frac{1}{2} \, \Big(\frac{1}{A} - \alpha\Big) 
\Big(l^d m^a + m^d l^a\Big)\bigg] \, A \, h_{ab}
\nonumber \\
&=&
{h^d}_b,
\end{eqnarray}
where in the last equality we used
$m^a h_{ab}= 0 = l^a h_{ab}$.
Also,

\begin{eqnarray}\label{82_4}
\nabla_c^{(q)} l_d^{(q)} 
&=&
\partial_c l_d 
- \bigg[\frac{1}{2} \, q^{al} 
\big(- \nabla_l q_{cd} + \nabla_c q_{dl} + \nabla_d q_{cl}\big)
+ {\Gamma^a}_{cd}\bigg] \, l_a
\nonumber \\
&=&
\nabla_c l_d
+ \frac{1}{2} \, \alpha \, l^l \,
\big(- \nabla_l q_{cd} + \nabla_c q_{dl} + \nabla_d q_{cl}\big),
\end{eqnarray}
where in the second equality we used
$q^{ad} \, l_a = \alpha \, l^d$.
From Eqs. (\ref{81_8}) and (\ref{82_4})
we get

\begin{eqnarray}\label{126_1}
K_{ab}^{(q)} 
&=& 
K_{ab} + \frac{1}{2} \, \alpha \, {h^c}_a {h^d}_b \, l^l \nabla_l q_{cd}
-\frac{1}{2} \, \alpha \, {h^c}_a {h^d}_b \, l^l 
(\nabla_c q_{dl} + \nabla_d q_{cl}),
\end{eqnarray}
which is the second step in equation (\ref{127_1}).

Now,

\begin{eqnarray}\label{83_1}
l^l \nabla_l q_{cd}
&=&
l^l \nabla_l \bigg[A \, g_{cd} - \frac{1}{2} \bigg(\frac{1}{\alpha} - A\bigg) 
(l_c m_d + m_c l_d)\bigg]
\nonumber \\
&=&
\frac{dA}{d\lambda} \, g_{cd} 
- \frac{1}{2} \, (l_c m_d + m_c l_d) \,
\frac{d}{d\lambda} \bigg(\frac{1}{\alpha} - A\bigg),
\nonumber
\end{eqnarray}
from
parallel transport along $\gamma$.
This gives

\begin{eqnarray}\label{83_3}
{h^c}_a {h^d}_b \, l^l \nabla_l q_{cd}
&=&
\frac{dA}{d\lambda} \, h_{ab}
\end{eqnarray}
(again using $m^a h_{ab}= 0 = l^a h_{ab}$). 
Further,

\begin{eqnarray}
l^l \nabla_c q_{dl} 
&=&
l_d \nabla_c A
- \frac{1}{2} \, l^l \nabla_c 
\bigg[\bigg(\frac{1}{\alpha} - A\bigg) \big(l_d m_l + m_d l_l\big)\bigg]
\nonumber \\
&=&
l_d \nabla_c A
+ l_d \nabla_c \bigg(\frac{1}{\alpha} - A\bigg)
- \frac{1}{2} \, \bigg(\frac{1}{\alpha} - A\bigg) \,
l^l \nabla_c \big(l_d m_l + m_d l_l\big)
\nonumber \\
&=&
l_d \nabla_c \frac{1}{\alpha}
- \frac{1}{2} \, \bigg(\frac{1}{\alpha} - A\bigg) \,
l^l \nabla_c \big(l_d m_l\big)
\nonumber \\
&=&
l_d \nabla_c \frac{1}{\alpha}
+ \bigg(\frac{1}{\alpha} - A\bigg) \, \nabla_c l_d
- \frac{1}{2} \, \bigg(\frac{1}{\alpha} - A\bigg) \,
l^l l_d \nabla_c m_l,
\nonumber
\end{eqnarray}
where we used $l^a m_a = -2$ and,
in the third step,
Eq. (\ref{100_3}).
From this,

\begin{eqnarray}\label{126_6}
{h^c}_a {h^d}_b \, l^l 
(\nabla_c q_{dl} + \nabla_d q_{cl})
&=&
{h^c}_a {h^d}_b \,
\bigg(\frac{1}{\alpha} - A\bigg) \,
\big(\nabla_c l_d + \nabla_d l_c\big)
\nonumber \\
&=&
\bigg(\frac{1}{\alpha} - A\bigg) \,
\big(K_{ab} + K_{ba}\big)
\nonumber \\
&=&
2 \, \bigg(\frac{1}{\alpha} - A\bigg) \, K_{ab},
\end{eqnarray}
where the last step is from the symmetry of $K_{ab}$ 
(from the hypersurface orthogonality of the congruence $l^a$).
Using Eqs. (\ref{83_3}) and (\ref{126_6}) in Eq. (\ref{126_1}),
we get the third equality in Eq. (\ref{127_1})
and from it immediately the fourth.

We consider now how the expression (\ref{132_3}) for 
${\bar K}_{ab}^{\prime (q)}$ comes about.
In

\begin{eqnarray}
{\bar K}_{ab}^{\prime (q)} &=&  
{{h^c}_a}^{(q)} {{h^d}_b}^{(q)} \, \nabla_c^{(q)} {m'}_d^{(q)},
\nonumber
\end{eqnarray}
the qmetric covariant derivative is given by

\begin{eqnarray}\label{128_1}
\nabla_c^{(q)} {m'}_d^{(q)}
&=&
\partial_c\bigg(\frac{1}{\alpha} \, m_d\bigg)
- {{\Gamma^a}_{cd}}^{(q)} \, \frac{1}{\alpha} \, m_a
\nonumber \\
&=&
\frac{1}{\alpha} \, \nabla_c m_d
+ m_d \, \partial_c \frac{1}{\alpha}
- \frac{1}{2} \, m^l \,
\big(- \nabla_l q_{cd} + \nabla_c q_{dl} + \nabla_d q_{cl}\big),
\nonumber
\end{eqnarray}
where in the first equality we used 
${m'}_c^{(q)} = (1/\alpha) \, m_c$,
and in the second, besides the connection (\ref{connection}), the relation
$q^{al} \, m_a = \alpha \, m^l$.
This gives

\begin{eqnarray}\label{128_3}
{\bar K}_{ab}^{\prime (q)} 
&=&
{{h^c}_a}^{(q)} {{h^d}_b}^{(q)} \,
\bigg[\frac{1}{\alpha} \, \nabla_c m_d
+ m_d \, \partial_c \frac{1}{\alpha}
- \frac{1}{2} \, m^l \,
\big(- \nabla_l q_{cd} + \nabla_c q_{dl} + \nabla_d q_{cl}\big)\bigg]
\nonumber \\
&=&
\frac{1}{\alpha} \, {\bar K}_{ab}
+ \frac{1}{2} \, {h^c}_a {h^d}_b \, m^l \nabla_l q_{cd}
-\frac{1}{2} \, {h^c}_a {h^d}_b \, m^l 
\big(\nabla_c q_{dl} + \nabla_d q_{cl}\big),
\end{eqnarray}
where we used Eq. (\ref{81_8}) and the fact that
${h^d}_b \, m_d = 0$ [or that ${h^c}_a \, \partial_c \frac{1}{\alpha} = 0$
since $\alpha = {\rm const}$ on $\Sigma(P, \lambda)$]. 

From

\begin{eqnarray}
m^l \nabla_l q_{cd} 
&=&
m^l \nabla_l \bigg[A \, g_{cd} - \frac{1}{2} \bigg(\frac{1}{\alpha} - A\bigg) 
(l_c m_d + m_c l_d)\bigg]
\nonumber \\
&=&
\frac{dA}{d\nu} \, g_{cd} 
- \frac{1}{2} \, (l_c m_d + m_c l_d) \,
\frac{d}{d\nu} \bigg(\frac{1}{\alpha} - A\bigg),
\nonumber
\end{eqnarray}
where we used Eq. (\ref{102_6})
as applied alternately to $z^a = l^a$ or $z^a = m^a$,
we get

\begin{eqnarray}\label{129_1}
{h^c}_a {h^d}_b \, m^l \nabla_l q_{cd}
&=&
\frac{dA}{d\nu} \,  h_{ab}
\end{eqnarray}
due to the orthogonality of $m^a$ or $l^a$ to $\Sigma(P, \lambda)$
[cf. Eq. (\ref{83_3})].
Also,

\begin{eqnarray}\label{129_2}
m^l \nabla_c q_{dl} 
&=&
m_d \nabla_c A 
+ m_d \nabla_c \bigg(\frac{1}{\alpha} - A\bigg)
- \frac{1}{2} \, \bigg(\frac{1}{\alpha} - A\bigg) \,
m^l \nabla_c \big(l_d m_l + m_d l_l\big)
\nonumber \\
&=&
m_d \nabla_c \frac{1}{\alpha}
- \frac{1}{2} \, \bigg(\frac{1}{\alpha} - A\bigg) \,
m^l \nabla_c \big(m_d l_l\big)
\nonumber \\
&=&
m_d \nabla_c \frac{1}{\alpha}
+ \bigg(\frac{1}{\alpha} - A\bigg) \, \nabla_c m_d
- \frac{1}{2} \, \bigg(\frac{1}{\alpha} - A\bigg) \,
m^l m_d \nabla_c l_l,
\nonumber
\end{eqnarray}
where in the second step we used Eq. (\ref{100_6}); 
this gives

\begin{eqnarray}\label{129_5}
{h^c}_a {h^d}_b \, m^l \, 
\big(\nabla_c q_{dl} + \nabla_d q_{cl}\big)
&=&
{h^c}_a {h^d}_b \, 
\bigg(\frac{1}{\alpha} - A\bigg)
\big(\nabla_c m_d + \nabla_d m_c\big)
\nonumber \\
&=&
2 \, \bigg(\frac{1}{\alpha} - A\bigg) \, {\bar K_{ab}},
\end{eqnarray}
where the first step comes from the orthogonality 
of $m^a$ to $\Sigma(P, \lambda)$,
and the second from the symmetry of $\bar K_{ab}$
[which is clear for example thinking to the geodesic congruence
of tangent $m^a$ on $\Sigma(P, \lambda)$, which is orthogonal
to the hypersurface $C \times \Sigma(P, \lambda)$  
at any $p \in L$, with $C$ any null curve with tangent $m^a$ at $p$]. 
Inserting Eqs. (\ref{129_1}) and (\ref{129_5}) into
Eq. (\ref{128_3}),
and using Eq. (\ref{133_0}) with $\Phi = A$,
we get

\begin{eqnarray}
{\bar K}_{ab}^{\prime (q)} 
&=&
A \, {\bar K_{ab}}
+ \frac{1}{2} \, \frac{dA}{d\nu} \, h_{ab}
\nonumber \\
&=&
 A \, {\bar K_{ab}}
- \frac{1}{2} \, \frac{dA}{d\lambda} \, h_{ab}, 
\nonumber
\end{eqnarray}
which is Eq. (\ref{132_3}).

Let us consider finally the quantity
$K^{ab}_{(q)} {\bar K}_{ab}^{\prime (q)}$
[of Eq. (\ref{K_barK})].  
To evaluate it, we need an expression for $K^{ab}_{(q)}$.
This is

\begin{eqnarray}
K^{ab}_{(q)}
&=&
q^{ac} q^{bd} K_{cd}^{(q)}
\nonumber \\
&=&
q^{ac} q^{bd} \,
\bigg(\alpha A \, K_{cd} 
+ \frac{1}{2} \, \alpha \, \frac{dA}{d\lambda} \, h_{cd}\bigg)
\nonumber \\
&=&
\frac{1}{A} \, g^{ac} \, \frac{1}{A} \, g^{bd} \,
\bigg(\alpha A \, K_{cd} 
+ \frac{1}{2} \, \alpha \, \frac{dA}{d\lambda} \, h_{cd}\bigg)
\nonumber \\
&=&
\frac{\alpha}{A} \, \bigg[K^{ab}
+ \frac{1}{2} \,
\bigg(\frac{d}{d\lambda} \ln A\bigg) \, h^{ab}\bigg],
\nonumber
\end{eqnarray}
where
we used Eq. (\ref{127_1})  in the second step,
and $l^a K_{ab} = 0 = m^a K_{ab}$ and $l^a h_{ab} = 0 = m^a h_{ab}$
in the third.
This gives

\begin{eqnarray}
K^{ab}_{(q)} {\bar K}_{ab}^{\prime (q)}
&=&
\frac{\alpha}{A} \, \bigg[K^{ab}
+ \frac{1}{2} \,
\bigg(\frac{d}{d\lambda} \ln A\bigg) \, h^{ab}\bigg] \,
\bigg(A \, {\bar K_{ab}}
- \frac{1}{2} \, \frac{dA}{d\lambda} \, h_{ab}\bigg)
\nonumber \\
&=&
\alpha \, K^{ab} {\bar K_{ab}}
+ \frac{1}{2} \, \alpha \, \Big(\frac{d}{d\lambda} \ln A\Big) 
({\bar K} - K)
- \frac{1}{4} (D-2) \, \alpha \, \Big(\frac{d}{d\lambda} \ln A\Big)^2,
\end{eqnarray}
where we used
$K^{ab} h_{ab} =
K_{ab} h^{ab} =
K_{ab} \big[g^{ab} + (1/2) \, l^a m^b + (1/2) \, m^a l^b\big] =
K_{ab} g^{ab} =
K
$
and, analogously,
${\bar K_{ab}} h^{ab} = \bar K$.

\section{Derivation of Eq. (\ref{162_1})}\label{AppD}

Concerning how
the expression for $dK/d\lambda$ [Eq. (\ref{162_1})] comes about,
we have (calculations are in $M^\sharp$, defined in Sec. II)

\begin{eqnarray}
\frac{dK}{d\lambda}
&=&
l^a \nabla_a K
\nonumber \\
&=&
l^a \nabla_a \nabla_b l^b
\nonumber \\
&=&
(l^a \nabla_b \nabla_a l^b)^\sharp
+ l^a {R^{\sharp b}}_{nab} l^n
\nonumber \\
&=&
(l^a \nabla_b \nabla_a l^b)^\sharp
- R_{ab}^\sharp l^a l^b
\nonumber \\
&=&
[\nabla_b (l^a \nabla_a l^b)]^\sharp
- \nabla_b  l^a \, \nabla_a l^b 
-  R_{ab}^\sharp l^a l^b,
\nonumber \\
&=&
- \nabla_b  l^a \, \nabla_a l^b 
-  R_{ab}^\sharp l^a l^b,
\end{eqnarray}
where
in the second equality we used Eq. (\ref{108_3}),
and in the third the relation 
($\nabla_d \nabla_c w^m - \nabla_c \nabla_d w^m)^\sharp
= - {R^{\sharp m}}_{ncd} w^n$ for any vector field $w^m$.
The last step
stems
from Eq. (\ref{102_6}) as applied
to the vector
(identically vanishing on $L$)
$z^b = l^a \nabla_a l^b$.

Moreover,

\begin{eqnarray}
\nabla_b  l^a \, \, \nabla_a l^b
&=&
g^{af} g^{bg} \, \nabla_b l_f \, \nabla_a l_g
\nonumber \\
&=&
\bigg(h^{af} - \frac{1}{2} \, l^a m^f - \frac{1}{2} \, m^a l^f\bigg)
\bigg(h^{bg} - \frac{1}{2} \, l^b m^g - \frac{1}{2} \, m^b l^g\bigg) \,
\nabla_b l_f \, \, \nabla_a l_g
\nonumber \\
&=&
h^{af} h^{bg} \, \nabla_b l_f \, \nabla_a l_g
\nonumber \\
&=&
h^{af} h^{bg} \, \nabla_f l_b \, \nabla_a l_g
\nonumber \\
&=&
{h^a}_l h^{lf} {h^b}_m h^{mg} \, \nabla_f l_b \, \nabla_a l_g
\nonumber \\
&=&
{K^l}_m {K_l}^m,
\end{eqnarray}
where the third equality comes from $l^a$ being
parallel displaced along $\gamma$ and from Eq. (\ref{100_3}) 
[or from Eq. (\ref{102_6}) with $z^b = l^b$],
the fourth from Eq. (\ref{72_2}) and  
$h^{ab} l_a = 0$,
and the fifth from ${h^a}_l h^{lf} = h^{af}$.
From this, we get Eq. (\ref{162_1}).

\section{Proof of Eq. (\ref{Q1_70_4})}\label{AppE}


From the relation connecting
Weyl ($C_{abcd}$) and Riemann tensors,

\begin{eqnarray}
C_{abcd} =
R_{abcd}
+ \frac{1}{D-2} \, 
\big(g_{ad} R_{cb} - g_{ac} R_{db} + g_{bc} R_{da} - g_{bd} R_{ca}\big)
+ \frac{1}{(D-1) (D-2)} \, R \, \big(g_{ac} g_{db} - g_{ad} g_{cb}\big), 
\nonumber
\end{eqnarray}
we get

\begin{eqnarray}
{h^c}_a {h^d}_b \, C_{cmdn} l^m l^n
&=&
{h^c}_a {h^d}_b \, R_{cmdn} l^m l^n
- h_{ab} R_{mn} l^m l^n
\nonumber \\
&=&
{h^c}_a {h^d}_b E_{cd} 
- h_{ab} R_{mn} l^m l^n.
\end{eqnarray}
Since we already know that $R_{ab}^\sharp l^a l^b = R_{ab} l^a l^b$ 
[Eq. (\ref{Riccill})],
we see that we have
${h^c}_a {h^d}_b E_{cd}^\sharp =
{h^c}_a {h^d}_b E_{cd}$ 
[i.e. Eq. (\ref{Q1_70_4})]
if and only if
${h^c}_a {h^d}_b \, C_{cmdn}^\sharp l^m l^n =
{h^c}_a {h^d}_b \, C_{cmdn} l^m l^n$.

The evolution equation 
of the shear
$\sigma_{ab}$ of our congruence
connects the Weyl tensor 
at the points along the geodesics
with the parameters characterizing the congruence. 
Using the (pseudo-orthonormal) basis
$\{l^a, \frac{1}{2} m^a, u^a_A\}$, $A = 1, ..., D-2$,
(where the $u^a_A$'s are unit spacelike and orthogonal 
to each other and to $l^a$ and $m^a$)
parallel transported along the congruence,
this evolution equation reads
[cf. Eq. (4.36) of \cite{HawB}]

\begin{eqnarray}\label{Q1_69_1}
\frac{d\sigma_{AB}}{d\lambda} =
- C_{abcd} \, u^a_A l^b u^c_B l^d
-\theta \, \sigma_{AB} - \sigma_{ac} {\sigma^c}_b \, u^a_A u^b_B
+ h_{AB} \, \sigma^{ab} \sigma_{ab},
\end{eqnarray} 
where we have written it for the rotation or vorticity 
tensor $\omega_{ab} = 0$, 
which is our case 
(from the hypersurface orthogonality of the $l^a$ vector field).
Here,
$B = 1, ..., D-2$ like $A$,
$\sigma_{AB} = \sigma_{ab} u^a_A u^b_B$,
$h_{AB} = h_{ab} u^a_A u^b_B$
and $\theta$ is expansion.

Equation (\ref{Q1_69_1}) shows that

\begin{eqnarray}\label{Q1_69_2}
C_{abcd}^\sharp \, u^a_A l^b u^c_B l^d =
C_{abcd} \, u^a_A l^b u^c_B l^d,
\end{eqnarray}
for all the terms entering Eq. (\ref{Q1_69_1})
beside the Weyl term
manifestly depend only on quantities 
which are assigned with the congruence
(and the latter is left untouched when going from $M$ to $M^\sharp$).  

Now,

\begin{eqnarray}
{h^c}_a =  
\sum_A u^c_A \, u_{Aa}.
\end{eqnarray}
This gives

\begin{eqnarray}
{h^c}_a {h^d}_b \, C_{cmdn} \, l^m l^n
&=&
\Big(\sum_A u^c_A \, u_{Aa}\Big) \Big(\sum_B u^d_B \, u_{Bb}\Big)
\, C_{cmdn} \, l^m l^n
\nonumber \\
&=&
\sum_{A, B} u_{Aa} u_{Bb} \, C_{cmdn} \, u^c_A l^m u^d_B l^n.
\end{eqnarray}
From Eq. (\ref{Q1_69_2}), this gives 

\begin{eqnarray}
{h^c}_a {h^d}_b \, C_{cmdn}^\sharp \, l^m l^n =
{h^c}_a {h^d}_b \, C_{cmdn} \, l^m l^n,
\end{eqnarray}
and then
${h^c}_a {h^d}_b E_{cd}^\sharp =
{h^c}_a {h^d}_b E_{cd}$,
which is Eq. (\ref{Q1_70_4}).

\section{Evaluation of the limits (\ref{epsilon})}\label{AppF}

Recall how we have introduced the vector field $V^a$.
It is obtained picking a local Lorentz frame at $P$, 
with its unit basis time vector ${\hat V^a} \equiv V^a_{|P}$,
and then parallel transporting it 
along each null geodesic $\gamma$.

This construction implies, in particular, that 
$V^a$ is parallel along any null $l^a$ at $P$,
and thus that $V^a$ is parallel along any direction at $P$,
i.e.

\begin{eqnarray}\label{nablaV}
\nabla_a V^b = 0,
\end{eqnarray}
at $P$.

But,

\begin{eqnarray}\label{epsilon_ab}
\epsilon_{ab} 
\equiv
K_{ab} + {\bar K_{ab}}
=
{h^c}_a {h^d}_b \, (\nabla_c l_d + \nabla_c m_d)
=
2 \, {h^c}_a {h^d}_b \, \nabla_c V_d,
\end{eqnarray}
at $p\in L$.
This gives 
$\epsilon_{ab} = 0$ at $P$
and, by continuity (of the last form of $\epsilon_{ab}$),
$\epsilon_{ab} \to 0$ for $\lambda \to 0$.
This establishes the first of the limits (\ref{epsilon}).

As for the second limit,
we notice that 

\begin{eqnarray}\label{reggeq_41_2}
\frac{1}{2} \, \frac{d\epsilon}{d\lambda}
&=&
l^a \, \nabla_a \nabla_b V^b
\nonumber \\
&=&
(l^a \, \nabla_b \nabla_a V^b)^\sharp
- l^a {R^{\sharp b}}_{dab} V^d
\nonumber \\
&=&
(\nabla_b(l^a\nabla_a V^b))^\sharp 
- (\nabla_b l^a) \nabla_a V^b
+ {R^{\sharp b}}_{abd} l^a V^d
\nonumber \\
&=&
- (\nabla_b l^a) \nabla_a V^b,
\end{eqnarray}
with $p\in L$.
Here, the last equality is from 
both
$(\nabla_b(l^a\nabla_a V^b))^\sharp = 0$
and
${R^{\sharp b}}_{abd} l^a V^d = 0$
at $p\in L$.
The first of these arises from the fact that
the vector field
$z^b \equiv l^a\nabla_a V^b = 0$ everywhere on $L$
and from the use of Eq. (\ref{133_0_bis}) for it, 
which implies 
$\nabla_c(l^a\nabla_a V^b) = 0$;
the second arises
from the relations
$
(V^a \nabla_a (l^b \nabla_b l^c))^\sharp
=
(V^a \nabla_a (m^b \nabla_b l^c))^\sharp
=
(V^a \nabla_a (e^b_A \nabla_b l^c))^\sharp
= 0
$
($p\in L$),
which give
$
{R^{\sharp c}}_{dba} l^d l^b V^a
=
{R^{\sharp c}}_{dba} l^d m^b V^a
=
{R^{\sharp c}}_{dba} l^d e^b_A V^a
=
0
$
[cf. Eq. (\ref{Q1_28_1})]
and then
$
{R^{\sharp c}}_{dba} l^d V^a = 0,
$
$p \in L$.
Since Eq. (\ref{reggeq_41_2}) holds in $M^\sharp$,
it holds in any deformation of $M^\sharp$ (and $M$)
which leaves $L$ unchanged,
for both sides of the equation are left unchanged
by any such deformation ($\epsilon_{ab}$ is left unchanged
since both $K_{ab}$ and $\bar K_{ab}$ do not change). 

Now,
we have just seen, right above, 
that
$\nabla_a V^b \to 0$ when $\lambda \to 0$; 
from Eq. (\ref{reggeq_41_2}) this gives 
$d\epsilon/d\lambda \to 0$ when $\lambda \to 0$,
which is the second limit (\ref{epsilon}).


\end{document}